\documentclass[aps,prl,amsmath,twocolumn,nopacs,groupedaddress]{revtex4-1}
\usepackage[english]{babel}
\usepackage{graphicx}
\usepackage{subfigure}
\usepackage{dcolumn}
\usepackage{bm}
\usepackage{units}
\usepackage{color}
\usepackage{longtable}
\usepackage{array}
\usepackage{multirow}
\usepackage{tablefootnote}
\usepackage[colorlinks=true,linkcolor=blue,citecolor=blue,urlcolor=blue,filecolor=blue]{hyperref}

    \renewcommand{\bottomfraction}{1}	
    \renewcommand{\textfraction}{0.07}	

\newcolumntype{L}[1]{>{\raggedright\let\newline\\\arraybackslash\hspace{0pt}}m{#1}}
\newcolumntype{C}[1]{>{\centering\let\newline\\\arraybackslash\hspace{0pt}}m{#1}}
\newcolumntype{R}[1]{>{\raggedleft\let\newline\\\arraybackslash\hspace{0pt}}m{#1}}

\begin{document}
\title{A steady-state magneto-optical trap with 100 fold improved phase-space density}

\author{Shayne Bennetts, Chun-Chia Chen, 
Benjamin Pasquiou}
\email[]{highPSDMOT@strontiumBEC.com}
\author{Florian Schreck}
\affiliation{Van der Waals-Zeeman Institute, Institute of Physics, University of Amsterdam, Science Park 904,
1098XH Amsterdam, The Netherlands}

\date{\today}

\begin{abstract}

We demonstrate a continuously loaded $^{88}\mathrm{Sr}$ magneto-optical trap (MOT) with a steady-state phase-space density of $1.3(2) \times 10^{-3}$. This is two orders of magnitude higher than reported in previous steady-state MOTs. Our approach is to flow atoms through a series of spatially separated laser cooling stages before capturing them in a MOT operated on the 7.4-kHz linewidth Sr intercombination line using a hybrid slower+MOT configuration. We also demonstrate producing a Bose-Einstein condensate at the MOT location, despite the presence of laser cooling light on resonance with the 30-MHz linewidth transition used to initially slow atoms in a separate chamber. Our steady-state high phase-space density MOT is an excellent starting point for a continuous atom laser and dead-time free atom interferometers or clocks.
\end{abstract}

\pacs{37.10.Gh, 37.10.De, 67.85.-d, 03.75.Be}


\maketitle

Laser cooled and trapped atoms are at the core of most ultracold quantum gas experiments \cite{Bloch2008ReviewManyBodyPhysics}, state-of-the-art clocks \cite{Ludlow2015oac} and sensors based on atom interferometry \cite{Cronin2009ReviewAtomInterferometry}. Today, these devices typically operate in a time-sequential manner, with distinct phases for sample preparation and measurement. For atomic clocks a consequence is the need to bridge the dead time between measurements using a secondary frequency reference, typically a resonator. This introduces a problem known as the Dick effect \cite{Dick1987loi} in which the sampling process inherent to a clock's cyclic operation down converts or aliases high frequency noise from the secondary reference into the signal band, thus degrading performance \cite{Westergaard2010DickEffect}. Recently, a new generation of atomic clocks using degenerate atoms in a three-dimensional optical lattice has been demonstrated using Sr \cite{Campbell2017afd}. To reach the potential of such a clock, it will be necessary to overcome the Dick effect, which can be achieved by reducing the dead time and/or by creating vastly improved secondary references. Our steady-state MOT can lead to significant advances in both directions. It approaches the high flux and low temperature requirements needed for a steady-state clock, which would completely eliminate the Dick effect. Furthermore, our MOT is created under conditions compatible with the creation of degenerate samples or an atom laser \cite{Stellmer2013LaserCoolingToBEC, Robins2013alp}. This would be the ideal source for a secondary frequency reference based on superradiant lasing, which is expected to outperform current references \cite{Chen2009aocActiveOpticalClock, Meiser2009MilliHzLaser, Meiser2010SteadyStateSuperradiance, Matei2017lws, Norcia2016SuperradianceStrontiumClock}. Our source and a future atom laser based on it might also be valuable for atomic inertial sensors \cite{Robins2013alp}. Improved clocks and inertial sensors will allow tests of fundamental physics \cite{Uzan2003FundamentalConstants} or be suitable for gravitational wave astronomy \cite{Kasevitch2013GravWave, Yu2011GWInterferometer, Ye2016GraveWaveClock, Cronin2009ReviewAtomInterferometry}.

Over the years many creative approaches have honed laser cooling to produce pulsed samples of ever increasing phase-space density (PSD) \cite{Lett1988SysiphusExp, Dalibard1989Sysiphus, Salomon2013K39GrayMollasses, Colzi2016NaGrayMolasses, Kasevich1992RamanCooling, Aspect1988VelocSelect, Ruhrig2015DemagnCoolingLimit, Hamann1998ResolvRamanSideband, Kerman2000RamanSideband3D, Han2000RamanSidebandCoolingWeiss, Yang2007ContinuousCaDTLoading, Radwell2013HighPSDDarkSPOT}. Pulsed MOTs using $^{88}\mathrm{Sr}$ have demonstrated phase-space densities of $10^{-2}$ \cite{Katori1999SrMOTRecoilTemp} while atoms held in dipole traps recently reached degeneracy \cite{Stellmer2013LaserCoolingToBEC, Hu2013LaserCoolingBEC}. Despite the exquisite performances, these techniques suffer from extremely small capture velocities. As a consequence atoms must first be captured and precooled, and thus these techniques have only been used as part of time-varying sequences.

Several continuous high PSD MOT schemes have been demonstrated mostly based on bichromatic MOTs using alkaline earth atoms \cite{Kawasaki2015TwoColorYbMOT, Lee2015CoreShellYbMOT}. The most successful reached a steady-state PSD of $1.2 \times 10^{-5}$ \cite{Hemmerich2002CaMOT}. This scheme used a MOT on a broad linewidth transition to capture atoms which then leak into a metastable state cooled by a MOT on a narrow transition. Narrow-line MOTs fed by a 2D MOT or Zeeman slower on the broad transition have been demonstrated for Yb and Er although steady-state PSDs are not measured \cite{Dorscher2013Yb2D3DMOT, Frisch2012ErNarrowLineMOT}. Another approach is the dark SPOT MOT \cite{Ketterle1993FirstDarkSPOT}, which creates a central spatial region of reduced laser interaction. Adding a further steady-state trapping and cooling stage to a MOT can increase the PSD substantially \cite{Mahnke2015ContinuousMagnReservoir, Falkenau2011ContinuousDTLoading, Volchkov2013sci}, with a steady-state PSD of $ \sim4 \times 10^{-4}$ reached for Cr atoms at a temperature of $\sim\unit[50]{\mu {\rm K}}$ \cite{Volchkov2013sci}.

\begin{figure}[tb]
\centering
\subfigure{\includegraphics[width=.98\columnwidth]{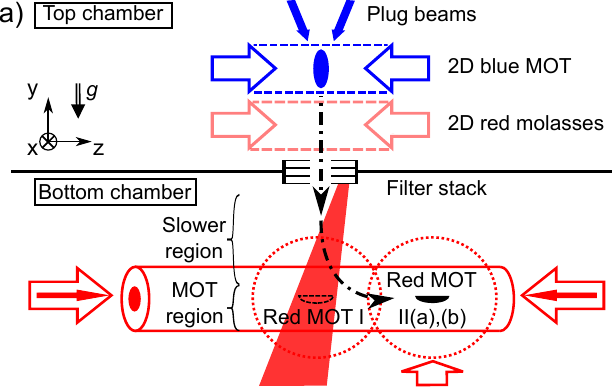}}
\subfigure{\includegraphics[width=.98\columnwidth]{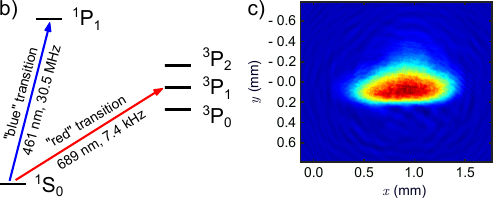}}
\caption{\label{fig:Fig1_ApparatusDesign} a) Schematic of our setup, showing the main cooling stages, and the position of the three ``red" MOT configurations (see text). b) Electronic level scheme of strontium, with the ``blue" and ``red" transitions used for laser cooling. c) In situ absorption picture of a steady-state $^{88}\mathrm{Sr}$ MOT with a PSD of $1.3(2) \times 10^{-3}$ (Red MOT II(b) configuration).}
\end{figure}

In this Letter, we demonstrate a $^{88}\mathrm{Sr}$ MOT at a temperature of $\sim\unit[2]{\mu {\rm K}}$ and a steady-state phase-space density of $1.3(2) \times 10^{-3}$, two orders of magnitude higher than reported for previous steady-state MOTs \cite{Hemmerich2002CaMOT}. Combining our MOT with techniques such as \cite{Chikkatur2002ContinuousBEC, Stellmer2013LaserCoolingToBEC, Robins2008PumpedAtomLaser, Mahnke2015ContinuousMagnReservoir, Falkenau2011ContinuousDTLoading, Volchkov2013sci} promises yet higher PSDs and potentially a steady-state BEC and atom laser. Our result is achieved by flowing atoms through a series of spatially separated cooling stages as illustrated in Figure~\ref{fig:Fig1_ApparatusDesign}. First we use the high capture velocity of the broad-linewidth, ``blue'', ${^1\mathrm{S}_0} - {^1\mathrm{P}_1}$ transition (\unit[30]{MHz} linewidth, \unit[461]{nm} wavelength) in several stages to slow and cool atoms to mK temperatures, finishing with a 2D ``blue'' MOT. Next, we capture the atoms in a 3D MOT using the narrow-linewidth, ``red", ${^1\mathrm{S}_0} - {^3\mathrm{P}_1}$ transition (\unit[7.4]{kHz} linewidth, \unit[689]{nm} wavelength), which can reach temperatures close to the recoil limit \cite{Katori1999SrMOTRecoilTemp}. We operate the two MOTs in separate chambers to avoid heating of the 3D ``red" MOT by blue photons scattered from surfaces and from fluorescing atoms in the 2D blue MOT. The transfer of atoms between chambers is ensured by two key ingredients: firstly, the atomic beam from the 2D blue MOT is collimated by a red optical molasses and secondly, the atoms are slowed in the second chamber by a hybrid slower+MOT configuration operated on the low capture velocity red transition. We show that this approach allows the red MOT to produce clouds with unprecedented phase-space densities for a steady-state apparatus. Furthermore, we show that the red MOT location is sufficiently protected to form BECs even with all the blue cooling stages operating.

The path of atoms through the setup begins with a high-flux atomic beam source adapted from our previous design \cite{StellmerPhD, Stellmer2009bec, SupplInf}. In brief, this source is composed of an oven similar to \cite{Senaratne2015OvenMicrotubes}, followed by a transverse cooling stage and a Zeeman slower, both using laser cooling on the broad-linewidth blue transition. A 2D blue MOT \cite{Dieckmann19982DMOT,Wilkowski20152DSrMOT,Weidemuller20172DSrMOT}, whose non-confining axis is oriented in the direction of gravity, is located approximately $\unit[5]{cm}$ after the exit of the Zeeman slower (see Figure~\ref{fig:Fig1_ApparatusDesign}a). This MOT has a loading rate of $\unit[2.66(16) \times 10^9]{^{88}Sr \, atoms / s}$ (measured by absorption imaging) and cools atoms to about $\unit[1]{mK}$ in the radial ($xz$) plane. To prevent atoms from escaping upward, a pair of downward propagating blue ``plug" beams are placed symmetrically to each other at an $\unit[8]{^\circ}$ angle from the $y$ axis, increasing the flux by around $\unit[30]{\%}$.

The high phase-space density MOT must be operated on the narrow-linewidth red transition while being protected against photons from the broad-linewidth blue transition \cite{Kawasaki2015TwoColorYbMOT}. In order to ensure such protection, we position the red MOT in a separate chamber, $\unit[41]{cm}$ below the 2D blue MOT. The two chambers are separated and baffled by a set of four stacked 1-inch absorptive neutral density filters (Thorlabs NE60A). These filters are separated by $\unit[20]{mm}$ and have an $\unit[8]{mm}$ diameter center hole allowing atoms to pass.

Upon exiting the 2D blue MOT, atoms have a radial velocity of about $\unit[0.5]{m/s}$, an average downward velocity of $\unit[3]{m/s}$ \cite{SupplInf}, and they fall accelerated by gravity towards the bottom chamber. Radial expansion during the $\unit[185]{mm}$ drop to the bottom of the baffle section would give a transfer efficiency less than $\unit[2]{\%}$. 
Accelerating the atoms downward is out of the question, since we can only use the low capture velocity red transition in the bottom chamber. Instead, we radially cool atoms emerging from the 2D MOT using a 2D molasses operating on the red ${^1\mathrm{S}_0} - {^3\mathrm{P}_1}$ $\pi$ transition, which is insensitive to the spatially varying magnetic field. To compensate the Doppler broadening of the atomic transition, which is more than 100 times the $\unit[7.4]{kHz}$ natural linewidth, it is necessary to modulate the frequency of the molasses laser beams forming a frequency comb spanning from $30$ to $\unit[750]{kHz}$ to the red of the transition, with $\unit[25]{kHz}$ spacing \cite{Stellmer2009bec}. This technique is used on all our red-transition laser beams. To prevent the two red ${^1\mathrm{S}_0} - {^3\mathrm{P}_1}$ $\sigma$ transitions towards the $m_{\mathrm{J'}} = \pm 1$ states from hindering the radial slowing process, we produce a Zeeman shift of about $\unit[3]{MHz}$ by applying a bias magnetic field of $\unit[1.4]{G}$ in the vertical direction. Such a small field doesn't disturb the operation of the 2D blue MOT. This molasses reaches steady state after a few $\unit[10]{ms}$, easily provided by four horizontal molasses beams with $1/e^2$ diameters of $\unit[45.6]{mm}$ along the $y$ axis.

After a $\unit[41]{cm}$ fall from the top to the bottom chamber, the atomic beam has a measured downward velocity distribution peaked at $\unit[4]{m/s}$. Our protection scheme necessitates slowing and capturing these falling atoms using only the red transition. However, the small scattering rate on this line allows a maximum deceleration of only $\sim 16 \, g$, where $g$ is earth acceleration. To overcome this extreme limitation we implement a hybrid slower+MOT.

Our first hybrid setup configuration labeled ``Red MOT I'' (see Figure~\ref{fig:Fig1_ApparatusDesign}a) uses a magnetic quadrupole field centered $\unit[23]{cm}$ directly below the bottom baffle between chambers. The gradients are $\unit[(0.55,0.35,0.23)]{G/cm}$ in the $(x,y,z)$ directions. Horizontally propagating laser beams in the $x,z$ axes are placed in a MOT configuration and provide radial cooling and confinement. On the $y$ axis, a single, upward propagating beam is used, with circular polarization as needed for the MOT. Due to the weakness of the transition this upward propagating beam and gravity are sufficient to confine atoms in the vertical direction without a downward propagating MOT beam \cite{Tey2010DoubleBECFermiSea}. The vertical beam is directed slightly to the side of the baffle onto the lowest neutral density filter, to prevent it from affecting the cooling processes in the top chamber. This beam is converging to exert a restoring force towards the beam center during the slowing process \cite{Prodan1982MonoEnergBeam}.

\begin{figure}[tb]
\centering
\subfigure{\includegraphics[width=.98\columnwidth]{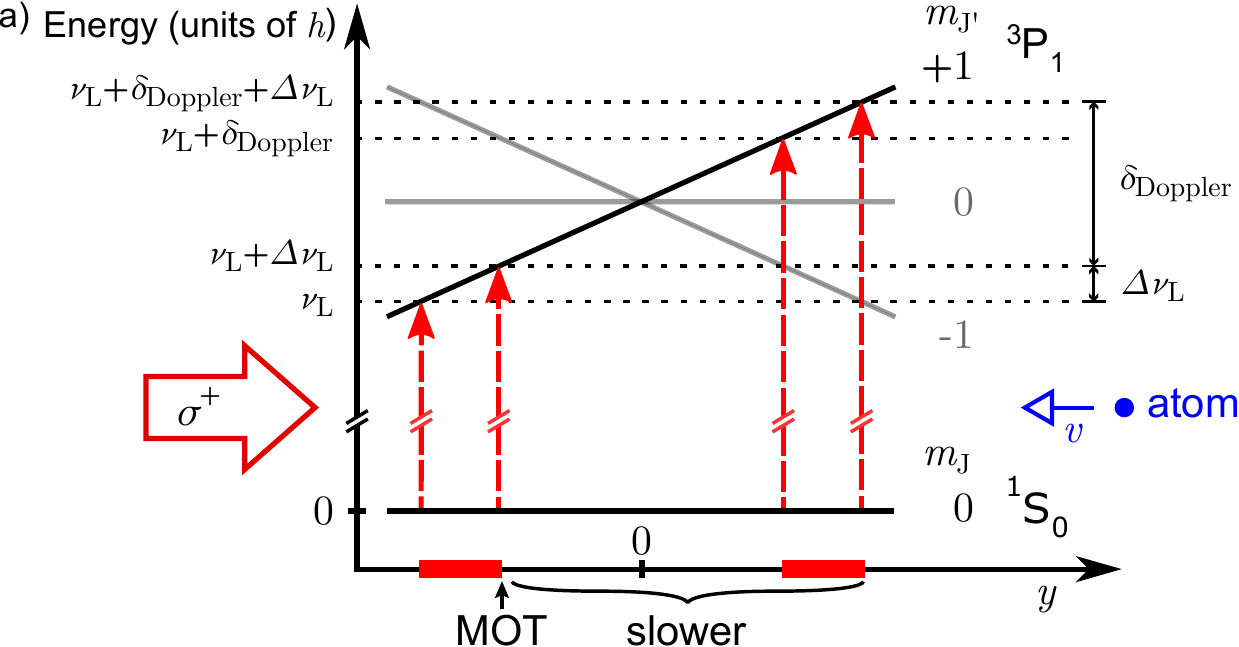}}
\subfigure{\includegraphics[width=.98\columnwidth]{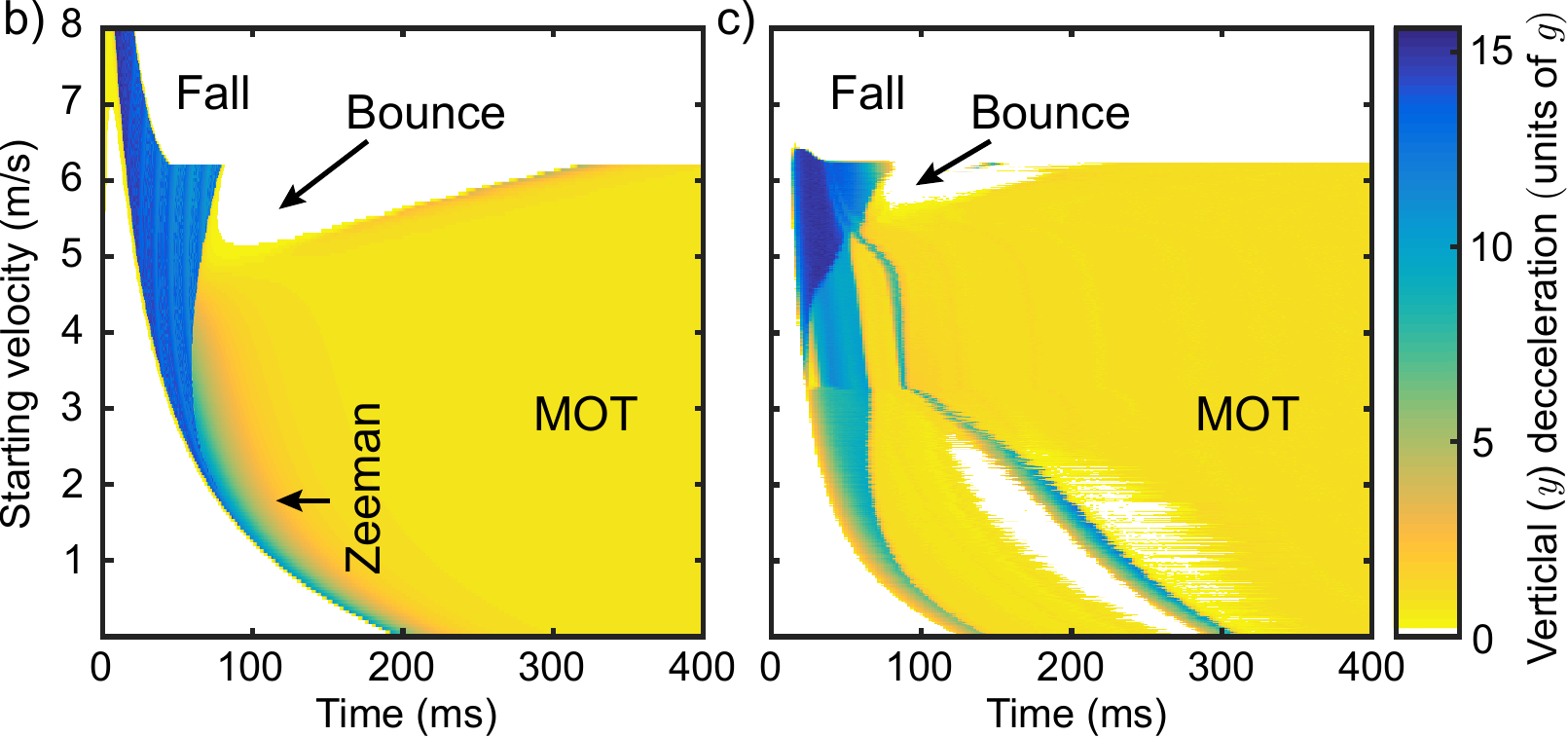}}
\caption{\label{fig:Fig2_HybridSlowerMOTPrinciple} Working principle of the hybrid slower+MOT in Red MOT I configuration. a) Energy diagram of the $^1\mathrm{S}_0 , m_{\mathrm{J}}$ and the $^3\mathrm{P}_1 , m_\mathrm{J'}$ states in dependence of $y$ for $x=z=0$, where the coordinate origin lays in the quadrupole field center. An atom with velocity $v \neq 0$ along the $y$ axis is addressed by the upward-propagating Red MOT I beam in a region delimited by the pair of red dashed arrows on the right. The range of this region is determined by the span of the laser frequency modulation $\Delta \nu_{\mathrm{L}}$ and the magnetic field gradient. An atom with $v = 0$ is addressed in the region delimited by the pair of dashed arrows on the left. b) $y$ deceleration from scattering of photons on the red transition (in units of earth acceleration $g$) in dependence of the initial velocity of atoms $\unit[20]{cm}$ above the quadrupole field center, obtained by an idealized 1D calculation. A horizontal path across the diagram corresponds to a 1D time-dependent trajectory along $y$. Annotations indicate regions dominated by Zeeman type slowing or red MOT type trapping. Atoms eventually captured in the MOT exhibit a time-independent non-zero force (yellow region at $\sim 1 g$). The white regions show decelerations smaller than $0.25 g$. The ``Bounce" and ``Fall" regions are undesirable cases described in \cite{SupplInf}. c) Equivalent deceleration data obtained by a full 3D Monte Carlo calculation with a realistic geometry.}
\end{figure}

Upon reaching the second chamber, atoms enter the region illuminated by the circularly-polarized upward-propagating beam, whose frequency is set to the red of the ${^1\mathrm{S}_0} - {^3\mathrm{P}_1}$ transitions. We now describe the slowing process using an upward pointing quantization axis, see Figure~\ref{fig:Fig2_HybridSlowerMOTPrinciple}a. The Doppler shift $\delta_{\mathrm{Doppler}} (v)$ of atoms with a downward velocity $v$ brings them into resonance with the $\sigma^+$ transition from $^1\mathrm{S}_0$ to the state $m_{\mathrm{J'}} = +1$ of $^3\mathrm{P}_1$. As radiation pressure slows atoms down, $\delta_{\mathrm{Doppler}}$ diminishes, which is partially compensated by the spatial variation of the magnetic field, following the principle of a Zeeman slower. The narrow linewidth of the red transition does not allow for a slowing process robust against magnetic and laser intensity fluctuations, so we modulate the laser frequency $\nu_{\mathrm{L}}$ with a span of $\Delta \nu_{\mathrm{L}} = \unit[4.05]{MHz}$, as used for example in ``white light" slowing \cite{Zhu1991FirstExpWhiteLightSlowing}. If atoms are successfully slowed and reach the region below the quadrupole field center, $\delta_{\mathrm{Doppler}}$ is small and atoms are resonant with the laser light as in a standard broadband narrow-line MOT \cite{Katori1999SrMOTRecoilTemp}. Note that for the experimental configurations used the angle between the local magnetic field and the beam direction can be big, leading to significant additional absorption on the $\pi$ transition during the slowing process.

We numerically model this hybrid setup by evolving classical atomic trajectories, first in an idealized 1D geometry with only a linear vertical magnetic field gradient and a uniform circularly polarized vertical beam, and then using a Monte Carlo approach in a realistic 3D geometry including all beams and details of the magnetic fields \cite{SupplInf}. The behavior of a falling atom can be obtained by analyzing the deceleration it experiences in dependence of time when dropped into the slower+MOT region with various starting velocities. The idealized 1D results shown in Figure~\ref{fig:Fig2_HybridSlowerMOTPrinciple}b are qualitatively confirmed by the more realistic 3D model results shown in Figure~\ref{fig:Fig2_HybridSlowerMOTPrinciple}c. With these simulations we estimate a maximum capture velocity of around $\unit[6]{m/s}$ for the hybrid slower+MOT setup, which is compatible with the measured velocity distribution of the atomic beam produced by the 2D blue MOT \cite{SupplInf}.

The characteristics of the hybrid Red MOT I are summarized in Table~\ref{tab:MOTParameters}. The loading rate gives an estimated transfer efficiency of $\unit[19]{\%}$ between the two chambers. Unfortunately, the high power broadband beams needed for the hybrid setup limit the PSD to $2.8(1.2) \times 10^{-6}$, low compared to what can be achieved with a pure red MOT geometry in a time-varying sequence \cite{Katori1999SrMOTRecoilTemp}. Moreover, with direct line-of-sight to a bright fluorescing blue MOT, this configuration does not provide protection against heating from blue photons sufficient to produce a steady-state BEC \cite{SupplInf}.

\begin{table}[tb]
\caption{Characteristics of the three Red MOT configurations. All uncertainties stated in this letter are taken as $\pm2\sigma$ from the fitted data.}
\begin{ruledtabular}
\begin{tabular}{C{0.13\textwidth} C{0.12\textwidth} C{0.11\textwidth} C{0.1\textwidth} }
 & \parbox[t]{0.09\textwidth}{Red MOT I} & \parbox[t]{0.09\textwidth}{Red MOT II(a)} & \parbox[t]{0.09\textwidth}{Red MOT II(b)}\\
 \noalign{\smallskip} \hline \noalign{\smallskip}
			Flux [$^{88}\mathrm{Sr/s}$] &  $5.1(7) \times 10^8$ & $5.3(4) \times 10^7$ & $5.3(9) \times 10^6$ \\
			Temperature $x$ & -\footnotemark[1] &  $\unit[3.7(3)]{\mu K}$ &  $\unit[2.01(6)]{\mu K}$\\
			Temperature $y$ & $\unit[20(5)]{\mu K}$  &  $\unit[1.9(1)]{\mu K}$ &  $\unit[1.42(3)]{\mu K}$\\
			Temperature $z$ & $\unit[26(7)]{\mu K}$  & $\unit[2.8(2)]{\mu K}$ & $\unit[1.91(9)]{\mu K}$ \\
			Width $\sigma_{x}$ & -\footnotemark[1]  & $\unit[385(4)]{\mu m}$ & $\unit[150(2)]{\mu m}$ \\
			Width $\sigma_{y}$ & $\unit[725(61)]{\mu m}$  &  $\unit[192(3)]{\mu m}$ &  $\unit[88(1)]{\mu m}$\\
			Width $\sigma_{z}$ & $\unit[2086(41)]{\mu m}$  & $\unit[528(3)]{\mu m}$ & $\unit[247(3)]{\mu m}$ \\
			Atom number [$^{88}\mathrm{Sr}$] & $2.54(10) \times 10^9$ & $1.71(5) \times 10^8$ & $2.5(1) \times 10^7$  \\
			Peak density [$^{88}\mathrm{Sr/cm^3}$]  & $5.1(7) \times 10^{10}$ & $2.8(2) \times 10^{11}$  & $4.8(4) \times 10^{11}$  \\
			Peak PSD & $2.8(1.2) \times 10^{-6}$ & $4(1) \times 10^{-4}$  & $1.3(2) \times 10^{-3}$  \\
			$1/e$ lifetime & $\unit[4.53(6)]{s}$ & $\unit[2.8(2)]{s}$  & $\unit[1.95(6)]{s}$  \\
\end{tabular}
\end{ruledtabular}
\footnotetext[1]{At the location of Red MOT I no imaging system for the $x$ axis was available so Red MOT I density and PSD calculations assume that  temperature and width are the same as z along x.}
\label{tab:MOTParameters}
\end{table}

For these reasons, we implement a second MOT configuration, ``Red MOT II(a)'', located $\unit[3]{cm}$ in the $z$ direction away from the Red MOT I position (see Figure~\ref{fig:Fig1_ApparatusDesign}a). This is achieved by adding another 5-beam geometry MOT (4 beams along $x$ and $z$, 1 beam along $y$), whose beams make a smooth connection with the Red MOT I beams, and by displacing the center of the quadrupole field to the intersection of these new beams. Along the $z$ axis the two additional beams are implemented as an $\unit[8]{mm}$ diameter low-intensity core within hollow $\unit[48]{mm}$ diameter Red MOT I beams. In this configuration, atoms entering the second chamber are decelerated by the hybrid slower plus Red MOT I beams, then continuously pushed to the Red MOT II(a) where they are confined and further cooled. The working principle of the hybrid setup remains unchanged, with the exception that the $\pi$ transition becomes more important, since the magnetic field lines are more tilted with respect to the Red MOT I vertical beam.

Within this second red MOT, we have more freedom to adapt the beam parameters and thereby reach higher PSD. The main limitations to the PSD are firstly power broadening of the red transition effectively raising the Doppler cooling limit and secondly multiple scattering due to the high density. Both limitations recede with reduced MOT powers. For this reason, we implement a spectrally dark SPOT MOT by shaping the spectral intensity profile of our broadband MOT beams. At the steady-state MOT location, the effective scattering rate is just enough to hold atoms against gravity, while in the surrounding region an effective high optical power captures fast atoms.

With Red MOT II(a) using the parameters given in Table~\ref{tab:MOTParameters}, we reach a steady-state MOT with a PSD of $4(1) \times 10^{-4}$ for $^{88}\mathrm{Sr}$. The transfer efficiency from the top chamber to this MOT is $\unit[2]{\%}$, an order of magnitude lower than for Red MOT I. Simulations suggest significant losses may be attributed firstly to increased bouncing of atoms in the hybrid slower when the $\pi$ transition is dominant and secondly to atom trajectories intersecting with mirrors inside the vacuum chamber.

We also produce a Red MOT II(a) using the much less abundant $^{84}\mathrm{Sr}$ isotope, which is particularly suited to produce quantum degenerate gases, owing to its favorable scattering properties. This $^{84}\mathrm{Sr}$ MOT contains up to $9.0(2) \times 10^{6}$ atoms at temperatures of $\unit[1.5(3)]{\mu K}$ and $\unit[3.4(1.1)]{\mu K}$ in the vertical and horizontal axes respectively. By loading this MOT into an optical dipole trap in a time sequential manner and applying a $\unit[3]{s}$ evaporative cooling sequence, we produce $^{84}\mathrm{Sr}$ BECs of $3.0(2) \times 10^{5}$ atoms.

The protection provided by the dual chamber and baffle system against photons from the broad-linewidth blue transition is determined by comparing the BEC lifetime at the location of the Red MOT II(a) with and without the 2D blue MOT operating. The atomic flux is disabled by turning off the red transition beams during these measurements. The BEC lifetime in the dipole trap is $\unit[3.0(4)]{s}$ without blue light and $\unit[2.7(2)]{s}$ with blue light. Lifetimes were strongly limited by one-body collisions due to poor vacuum quality. These measurements confirm significant protection from blue photons thanks to our two-chamber design, making our system suitable for experiments aimed at developing a steady-state source of degenerate quantum gas.

Finally, ``Red MOT II(b)" is a configuration designed to optimize phase-space density at the expense of transfer efficiency. This MOT reaches a steady-state PSD of $1.3(2) \times 10^{-3}$ for $^{88}\mathrm{Sr}$, which is two orders of magnitude higher than demonstrated in previous steady-state MOTs \cite{Hemmerich2002CaMOT} (see Figure~\ref{fig:Fig1_ApparatusDesign}c and Table~\ref{tab:MOTParameters}). Red MOT II(b) is the same as Red MOT II(a) except that we reduce the bandwidth of the MOT II beams to cover only $-40$ to $\unit[-200]{kHz}$ detuning, resulting in three improvements. First, the smaller detuning of $\unit[-40]{kHz}$ compresses the MOT, second, the reduced bandwidth reduces the total beam intensity required, and third, by ending the spectrum well before the photoassociation line located at $\unit[-435]{kHz}$ detuning \cite{Borkowski2014SrMassScalingPot}, we reduce losses due to molecule formation. The transfer efficiency from the top chamber to this MOT is $\unit[0.2]{\%}$, an order of magnitude lower than for Red MOT II(a). The $1/e$ lifetime of Red MOT II(b) measured after the atomic flux is suddenly stopped is $\unit[1.95(6)]{s}$, which is significantly smaller than the $\unit[4.53(6)]{s}$ lifetime of Red MOT I. This reduction is due to two-body light-assisted collisions, which ultimately limits the maximum density achievable \cite{Borkowski2014SrMassScalingPot}. Both these losses and the need to compensate for gravity set the limits on the PSD achievable with this MOT configuration.

To summarize, we have demonstrated the operation of a MOT with a PSD of $1.3(2) \times 10^{-3}$ in the steady-state regime. This result requires the use of broad- and narrow-linewidth transitions to simultaneously achieve both high PSD and high capture efficiency. We use a dual chamber architecture to protect the MOT from heating by broad-linewidth transition photons, and efficient transfer between chambers is achieved using a hybrid slower+MOT configuration using only cooling on the narrow-linewidth transition. Although Sr is ideally suited to this architecture, our approach is broadly applicable to alkaline-earth metals, lanthanides and any other species with a strong transition to precool atoms and a weak transition supporting the operation of a MOT with a high PSD \cite{Berglund2008NarrowLineErbiumMOT, Kuwamoto1999NarrowLineMOTYb}. Finally, we have shown that our design is compatible with the generation of quantum degenerate gases in the presence of a laser cooled influx. The use of this high-PSD source of matter, combined with a protection mechanism such as that demonstrated in \cite{Stellmer2013LaserCoolingToBEC}, should allow the creation of a steady-state Bose-Einstein condensate and ultimately an atom laser with uninterrupted phase coherent output.

\begin{acknowledgments}
We thank Georgios Siviloglou for contributions during the early stages of the design and construction of the experiment. We thank the Netherlands Organisation for Scientific Research (NWO) for funding through Vici grant No. 680-47-619. This project has received funding from the European Research Council (ERC) under the European Union's Seventh Framework Programme (FP7/2007-2013) (Grant agreement No. 615117 QuantStro). B.P. thanks the NWO for funding through Veni grant No. 680-47-438 and C.-C. C. thanks the Ministry of Education of the Republic of China (Taiwan) for a MOE Technologies Incubation Scholarship. S.B. thanks the Australian Government and Nick Robins at the Australian National University for an Australian Postgraduate Award and support during the early stages of this work.

S.B. and C.-C. C. contributed equally to this work.
\end{acknowledgments}



\begin{thebibliography}{55}%
\makeatletter
\providecommand \@ifxundefined [1]{%
 \@ifx{#1\undefined}
}%
\providecommand \@ifnum [1]{%
 \ifnum #1\expandafter \@firstoftwo
 \else \expandafter \@secondoftwo
 \fi
}%
\providecommand \@ifx [1]{%
 \ifx #1\expandafter \@firstoftwo
 \else \expandafter \@secondoftwo
 \fi
}%
\providecommand \natexlab [1]{#1}%
\providecommand \enquote  [1]{``#1''}%
\providecommand \bibnamefont  [1]{#1}%
\providecommand \bibfnamefont [1]{#1}%
\providecommand \citenamefont [1]{#1}%
\providecommand \href@noop [0]{\@secondoftwo}%
\providecommand \href [0]{\begingroup \@sanitize@url \@href}%
\providecommand \@href[1]{\@@startlink{#1}\@@href}%
\providecommand \@@href[1]{\endgroup#1\@@endlink}%
\providecommand \@sanitize@url [0]{\catcode `\\12\catcode `\$12\catcode
  `\&12\catcode `\#12\catcode `\^12\catcode `\_12\catcode `\%12\relax}%
\providecommand \@@startlink[1]{}%
\providecommand \@@endlink[0]{}%
\providecommand \url  [0]{\begingroup\@sanitize@url \@url }%
\providecommand \@url [1]{\endgroup\@href {#1}{\urlprefix }}%
\providecommand \urlprefix  [0]{URL }%
\providecommand \Eprint [0]{\href }%
\providecommand \doibase [0]{http://dx.doi.org/}%
\providecommand \selectlanguage [0]{\@gobble}%
\providecommand \bibinfo  [0]{\@secondoftwo}%
\providecommand \bibfield  [0]{\@secondoftwo}%
\providecommand \translation [1]{[#1]}%
\providecommand \BibitemOpen [0]{}%
\providecommand \bibitemStop [0]{}%
\providecommand \bibitemNoStop [0]{.\EOS\space}%
\providecommand \EOS [0]{\spacefactor3000\relax}%
\providecommand \BibitemShut  [1]{\csname bibitem#1\endcsname}%
\let\auto@bib@innerbib\@empty
\bibitem [{\citenamefont {Bloch}\ \emph {et~al.}(2008)\citenamefont {Bloch},
  \citenamefont {Dalibard},\ and\ \citenamefont
  {Zwerger}}]{Bloch2008ReviewManyBodyPhysics}%
  \BibitemOpen
  \bibfield  {author} {\bibinfo {author} {\bibfnamefont {I.}~\bibnamefont
  {Bloch}}, \bibinfo {author} {\bibfnamefont {J.}~\bibnamefont {Dalibard}}, \
  and\ \bibinfo {author} {\bibfnamefont {W.}~\bibnamefont {Zwerger}},\ }\href
  {\doibase 10.1103/RevModPhys.80.885} {\bibfield  {journal} {\bibinfo
  {journal} {Rev. Mod. Phys.}\ }\textbf {\bibinfo {volume} {80}},\ \bibinfo
  {pages} {885} (\bibinfo {year} {2008})}\BibitemShut {NoStop}%
\bibitem [{\citenamefont {Ludlow}\ \emph {et~al.}(2015)\citenamefont {Ludlow},
  \citenamefont {Boyd}, \citenamefont {Ye}, \citenamefont {Peik},\ and\
  \citenamefont {Schmidt}}]{Ludlow2015oac}%
  \BibitemOpen
  \bibfield  {author} {\bibinfo {author} {\bibfnamefont {A.~D.}\ \bibnamefont
  {Ludlow}}, \bibinfo {author} {\bibfnamefont {M.~M.}\ \bibnamefont {Boyd}},
  \bibinfo {author} {\bibfnamefont {J.}~\bibnamefont {Ye}}, \bibinfo {author}
  {\bibfnamefont {E.}~\bibnamefont {Peik}}, \ and\ \bibinfo {author}
  {\bibfnamefont {P.~O.}\ \bibnamefont {Schmidt}},\ }\href {\doibase
  10.1103/RevModPhys.87.637} {\bibfield  {journal} {\bibinfo  {journal} {Rev.
  Mod. Phys.}\ }\textbf {\bibinfo {volume} {87}},\ \bibinfo {pages} {637}
  (\bibinfo {year} {2015})}\BibitemShut {NoStop}%
\bibitem [{\citenamefont {Cronin}\ \emph {et~al.}(2009)\citenamefont {Cronin},
  \citenamefont {Schmiedmayer},\ and\ \citenamefont
  {Pritchard}}]{Cronin2009ReviewAtomInterferometry}%
  \BibitemOpen
  \bibfield  {author} {\bibinfo {author} {\bibfnamefont {A.~D.}\ \bibnamefont
  {Cronin}}, \bibinfo {author} {\bibfnamefont {J.}~\bibnamefont
  {Schmiedmayer}}, \ and\ \bibinfo {author} {\bibfnamefont {D.~E.}\
  \bibnamefont {Pritchard}},\ }\href {\doibase 10.1103/RevModPhys.81.1051}
  {\bibfield  {journal} {\bibinfo  {journal} {Rev. Mod. Phys.}\ }\textbf
  {\bibinfo {volume} {81}},\ \bibinfo {pages} {1051} (\bibinfo {year}
  {2009})}\BibitemShut {NoStop}%
\bibitem [{\citenamefont {Dick}(1987)}]{Dick1987loi}%
  \BibitemOpen
  \bibfield  {author} {\bibinfo {author} {\bibfnamefont {G.~J.}\ \bibnamefont
  {Dick}},\ }in\ \href
  {http://tycho.usno.navy.mil/ptti/1987papers/Vol\%2019_13.pdf} {\emph
  {\bibinfo {booktitle} {Proceedings of the Precise Time and Time Interval
  Meeting}}},\ \bibinfo {editor} {edited by\ \bibinfo {editor} {\bibfnamefont
  {R.}~\bibnamefont {Sydnow}}}\ (\bibinfo  {publisher} {U.S. Naval Observatory,
  Washington, DC},\ \bibinfo {year} {1987})\ p.\ \bibinfo {pages}
  {133}\BibitemShut {NoStop}%
\bibitem [{\citenamefont {Westergaard}\ \emph {et~al.}(2010)\citenamefont
  {Westergaard}, \citenamefont {Lodewyck},\ and\ \citenamefont
  {Lemonde}}]{Westergaard2010DickEffect}%
  \BibitemOpen
  \bibfield  {author} {\bibinfo {author} {\bibfnamefont {P.~G.}\ \bibnamefont
  {Westergaard}}, \bibinfo {author} {\bibfnamefont {J.}~\bibnamefont
  {Lodewyck}}, \ and\ \bibinfo {author} {\bibfnamefont {P.}~\bibnamefont
  {Lemonde}},\ }\href {\doibase 10.1109/TUFFC.2010.1457} {\bibfield  {journal}
  {\bibinfo  {journal} {IEEE Trans. Ultrason. Ferroelectr. Freq. Control}\
  }\textbf {\bibinfo {volume} {57}},\ \bibinfo {pages} {623} (\bibinfo {year}
  {2010})}\BibitemShut {NoStop}%
\bibitem [{\citenamefont {Campbell}\ \emph {et~al.}(2017)\citenamefont
  {Campbell}, \citenamefont {Hutson}, \citenamefont {Marti}, \citenamefont
  {Goban}, \citenamefont {Oppong}, \citenamefont {McNally}, \citenamefont
  {Sonderhouse}, \citenamefont {Robinson}, \citenamefont {Zhang}, \citenamefont
  {Bloom},\ and\ \citenamefont {Ye}}]{Campbell2017afd}%
  \BibitemOpen
  \bibfield  {author} {\bibinfo {author} {\bibfnamefont {S.~L.}\ \bibnamefont
  {Campbell}}, \bibinfo {author} {\bibfnamefont {R.~B.}\ \bibnamefont
  {Hutson}}, \bibinfo {author} {\bibfnamefont {G.~E.}\ \bibnamefont {Marti}},
  \bibinfo {author} {\bibfnamefont {A.}~\bibnamefont {Goban}}, \bibinfo
  {author} {\bibfnamefont {N.~D.}\ \bibnamefont {Oppong}}, \bibinfo {author}
  {\bibfnamefont {R.~L.}\ \bibnamefont {McNally}}, \bibinfo {author}
  {\bibfnamefont {L.}~\bibnamefont {Sonderhouse}}, \bibinfo {author}
  {\bibfnamefont {J.~M.}\ \bibnamefont {Robinson}}, \bibinfo {author}
  {\bibfnamefont {W.}~\bibnamefont {Zhang}}, \bibinfo {author} {\bibfnamefont
  {B.~J.}\ \bibnamefont {Bloom}}, \ and\ \bibinfo {author} {\bibfnamefont
  {J.}~\bibnamefont {Ye}},\ }\href {\doibase 10.1126/science.aam5538}
  {\bibfield  {journal} {\bibinfo  {journal} {Science}\ }\textbf {\bibinfo
  {volume} {358}},\ \bibinfo {pages} {90} (\bibinfo {year} {2017})}\BibitemShut
  {NoStop}%
\bibitem [{\citenamefont {Stellmer}\ \emph {et~al.}(2013)\citenamefont
  {Stellmer}, \citenamefont {Pasquiou}, \citenamefont {Grimm},\ and\
  \citenamefont {Schreck}}]{Stellmer2013LaserCoolingToBEC}%
  \BibitemOpen
  \bibfield  {author} {\bibinfo {author} {\bibfnamefont {S.}~\bibnamefont
  {Stellmer}}, \bibinfo {author} {\bibfnamefont {B.}~\bibnamefont {Pasquiou}},
  \bibinfo {author} {\bibfnamefont {R.}~\bibnamefont {Grimm}}, \ and\ \bibinfo
  {author} {\bibfnamefont {F.}~\bibnamefont {Schreck}},\ }\href {\doibase
  10.1103/PhysRevLett.110.263003} {\bibfield  {journal} {\bibinfo  {journal}
  {Phys. Rev. Lett.}\ }\textbf {\bibinfo {volume} {110}},\ \bibinfo {pages}
  {263003} (\bibinfo {year} {2013})}\BibitemShut {NoStop}%
\bibitem [{\citenamefont {Robins}\ \emph {et~al.}(2013)\citenamefont {Robins},
  \citenamefont {Altin}, \citenamefont {Debs},\ and\ \citenamefont
  {Close}}]{Robins2013alp}%
  \BibitemOpen
  \bibfield  {author} {\bibinfo {author} {\bibfnamefont {N.~P.}\ \bibnamefont
  {Robins}}, \bibinfo {author} {\bibfnamefont {P.~A.}\ \bibnamefont {Altin}},
  \bibinfo {author} {\bibfnamefont {J.~E.}\ \bibnamefont {Debs}}, \ and\
  \bibinfo {author} {\bibfnamefont {J.~D.}\ \bibnamefont {Close}},\ }\href
  {\doibase 10.1016/j.physrep.2013.03.006} {\bibfield  {journal} {\bibinfo
  {journal} {Phys. Rep.}\ }\textbf {\bibinfo {volume} {529}},\ \bibinfo {pages}
  {637} (\bibinfo {year} {2013})}\BibitemShut {NoStop}%
\bibitem [{\citenamefont {Chen}(2009)}]{Chen2009aocActiveOpticalClock}%
  \BibitemOpen
  \bibfield  {author} {\bibinfo {author} {\bibfnamefont {J.}~\bibnamefont
  {Chen}},\ }\href@noop {} {\bibfield  {journal} {\bibinfo  {journal} {Chin.
  Sci. Bull.}\ }\textbf {\bibinfo {volume} {54}},\ \bibinfo {pages} {348}
  (\bibinfo {year} {2009})}\BibitemShut {NoStop}%
\bibitem [{\citenamefont {Meiser}\ \emph {et~al.}(2009)\citenamefont {Meiser},
  \citenamefont {Ye}, \citenamefont {Carlson},\ and\ \citenamefont
  {Holland}}]{Meiser2009MilliHzLaser}%
  \BibitemOpen
  \bibfield  {author} {\bibinfo {author} {\bibfnamefont {D.}~\bibnamefont
  {Meiser}}, \bibinfo {author} {\bibfnamefont {J.}~\bibnamefont {Ye}}, \bibinfo
  {author} {\bibfnamefont {D.~R.}\ \bibnamefont {Carlson}}, \ and\ \bibinfo
  {author} {\bibfnamefont {M.~J.}\ \bibnamefont {Holland}},\ }\href {\doibase
  10.1103/PhysRevLett.102.163601} {\bibfield  {journal} {\bibinfo  {journal}
  {Phys. Rev. Lett.}\ }\textbf {\bibinfo {volume} {102}},\ \bibinfo {pages}
  {163601} (\bibinfo {year} {2009})}\BibitemShut {NoStop}%
\bibitem [{\citenamefont {Meiser}\ and\ \citenamefont
  {Holland}(2010)}]{Meiser2010SteadyStateSuperradiance}%
  \BibitemOpen
  \bibfield  {author} {\bibinfo {author} {\bibfnamefont {D.}~\bibnamefont
  {Meiser}}\ and\ \bibinfo {author} {\bibfnamefont {M.~J.}\ \bibnamefont
  {Holland}},\ }\href {\doibase 10.1103/PhysRevA.81.033847} {\bibfield
  {journal} {\bibinfo  {journal} {Phys. Rev. A}\ }\textbf {\bibinfo {volume}
  {81}},\ \bibinfo {pages} {033847} (\bibinfo {year} {2010})}\BibitemShut
  {NoStop}%
\bibitem [{\citenamefont {Matei}\ \emph {et~al.}(2017)\citenamefont {Matei},
  \citenamefont {Legero}, \citenamefont {H\"afner}, \citenamefont {Grebing},
  \citenamefont {Weyrich}, \citenamefont {Zhang}, \citenamefont {Sonderhouse},
  \citenamefont {Robinson}, \citenamefont {Ye}, \citenamefont {Riehle},\ and\
  \citenamefont {Sterr}}]{Matei2017lws}%
  \BibitemOpen
  \bibfield  {author} {\bibinfo {author} {\bibfnamefont {D.~G.}\ \bibnamefont
  {Matei}}, \bibinfo {author} {\bibfnamefont {T.}~\bibnamefont {Legero}},
  \bibinfo {author} {\bibfnamefont {S.}~\bibnamefont {H\"afner}}, \bibinfo
  {author} {\bibfnamefont {C.}~\bibnamefont {Grebing}}, \bibinfo {author}
  {\bibfnamefont {R.}~\bibnamefont {Weyrich}}, \bibinfo {author} {\bibfnamefont
  {W.}~\bibnamefont {Zhang}}, \bibinfo {author} {\bibfnamefont
  {L.}~\bibnamefont {Sonderhouse}}, \bibinfo {author} {\bibfnamefont {J.~M.}\
  \bibnamefont {Robinson}}, \bibinfo {author} {\bibfnamefont {J.}~\bibnamefont
  {Ye}}, \bibinfo {author} {\bibfnamefont {F.}~\bibnamefont {Riehle}}, \ and\
  \bibinfo {author} {\bibfnamefont {U.}~\bibnamefont {Sterr}},\ }\href
  {\doibase 10.1103/PhysRevLett.118.263202} {\bibfield  {journal} {\bibinfo
  {journal} {Phys. Rev. Lett.}\ }\textbf {\bibinfo {volume} {118}},\ \bibinfo
  {pages} {263202} (\bibinfo {year} {2017})}\BibitemShut {NoStop}%
\bibitem [{\citenamefont {Norcia}\ \emph {et~al.}(2016)\citenamefont {Norcia},
  \citenamefont {Winchester}, \citenamefont {Cline},\ and\ \citenamefont
  {Thompson}}]{Norcia2016SuperradianceStrontiumClock}%
  \BibitemOpen
  \bibfield  {author} {\bibinfo {author} {\bibfnamefont {M.~A.}\ \bibnamefont
  {Norcia}}, \bibinfo {author} {\bibfnamefont {M.~N.}\ \bibnamefont
  {Winchester}}, \bibinfo {author} {\bibfnamefont {J.~R.~K.}\ \bibnamefont
  {Cline}}, \ and\ \bibinfo {author} {\bibfnamefont {J.~K.}\ \bibnamefont
  {Thompson}},\ }\href {\doibase 10.1126/sciadv.1601231} {\bibfield  {journal}
  {\bibinfo  {journal} {Sci. Adv.}\ }\textbf {\bibinfo {volume} {2}},\ \bibinfo
  {pages} {e1601231} (\bibinfo {year} {2016})}\BibitemShut {NoStop}%
\bibitem [{\citenamefont {Uzan}(2003)}]{Uzan2003FundamentalConstants}%
  \BibitemOpen
  \bibfield  {author} {\bibinfo {author} {\bibfnamefont {J.-P.}\ \bibnamefont
  {Uzan}},\ }\href {\doibase 10.1103/RevModPhys.75.403} {\bibfield  {journal}
  {\bibinfo  {journal} {Rev. Mod. Phys.}\ }\textbf {\bibinfo {volume} {75}},\
  \bibinfo {pages} {403} (\bibinfo {year} {2003})}\BibitemShut {NoStop}%
\bibitem [{\citenamefont {Graham}\ \emph {et~al.}(2013)\citenamefont {Graham},
  \citenamefont {Hogan}, \citenamefont {Kasevich},\ and\ \citenamefont
  {Rajendran}}]{Kasevitch2013GravWave}%
  \BibitemOpen
  \bibfield  {author} {\bibinfo {author} {\bibfnamefont {P.~W.}\ \bibnamefont
  {Graham}}, \bibinfo {author} {\bibfnamefont {J.~M.}\ \bibnamefont {Hogan}},
  \bibinfo {author} {\bibfnamefont {M.~A.}\ \bibnamefont {Kasevich}}, \ and\
  \bibinfo {author} {\bibfnamefont {S.}~\bibnamefont {Rajendran}},\ }\href
  {\doibase 10.1103/PhysRevLett.110.171102} {\bibfield  {journal} {\bibinfo
  {journal} {Phys. Rev. Lett.}\ }\textbf {\bibinfo {volume} {110}},\ \bibinfo
  {pages} {171102} (\bibinfo {year} {2013})}\BibitemShut {NoStop}%
\bibitem [{\citenamefont {Yu}\ and\ \citenamefont
  {Tinto}(2011)}]{Yu2011GWInterferometer}%
  \BibitemOpen
  \bibfield  {author} {\bibinfo {author} {\bibfnamefont {N.}~\bibnamefont
  {Yu}}\ and\ \bibinfo {author} {\bibfnamefont {M.}~\bibnamefont {Tinto}},\
  }\href {\doibase 10.1007/s10714-010-1055-8} {\bibfield  {journal} {\bibinfo
  {journal} {Gen. Relat. and Gravit.}\ }\textbf {\bibinfo {volume} {43}},\
  \bibinfo {pages} {1943} (\bibinfo {year} {2011})}\BibitemShut {NoStop}%
\bibitem [{\citenamefont {Kolkowitz}\ \emph {et~al.}(2016)\citenamefont
  {Kolkowitz}, \citenamefont {Pikovski}, \citenamefont {Langellier},
  \citenamefont {Lukin}, \citenamefont {Walsworth},\ and\ \citenamefont
  {Ye}}]{Ye2016GraveWaveClock}%
  \BibitemOpen
  \bibfield  {author} {\bibinfo {author} {\bibfnamefont {S.}~\bibnamefont
  {Kolkowitz}}, \bibinfo {author} {\bibfnamefont {I.}~\bibnamefont {Pikovski}},
  \bibinfo {author} {\bibfnamefont {N.}~\bibnamefont {Langellier}}, \bibinfo
  {author} {\bibfnamefont {M.~D.}\ \bibnamefont {Lukin}}, \bibinfo {author}
  {\bibfnamefont {R.~L.}\ \bibnamefont {Walsworth}}, \ and\ \bibinfo {author}
  {\bibfnamefont {J.}~\bibnamefont {Ye}},\ }\href {\doibase
  10.1103/PhysRevD.94.124043} {\bibfield  {journal} {\bibinfo  {journal} {Phys.
  Rev. D}\ }\textbf {\bibinfo {volume} {94}},\ \bibinfo {pages} {124043}
  (\bibinfo {year} {2016})}\BibitemShut {NoStop}%
\bibitem [{\citenamefont {Lett}\ \emph {et~al.}(1988)\citenamefont {Lett},
  \citenamefont {Watts}, \citenamefont {Westbrook}, \citenamefont {Phillips},
  \citenamefont {Gould},\ and\ \citenamefont {Metcalf}}]{Lett1988SysiphusExp}%
  \BibitemOpen
  \bibfield  {author} {\bibinfo {author} {\bibfnamefont {P.~D.}\ \bibnamefont
  {Lett}}, \bibinfo {author} {\bibfnamefont {R.~N.}\ \bibnamefont {Watts}},
  \bibinfo {author} {\bibfnamefont {C.~I.}\ \bibnamefont {Westbrook}}, \bibinfo
  {author} {\bibfnamefont {W.~D.}\ \bibnamefont {Phillips}}, \bibinfo {author}
  {\bibfnamefont {P.~L.}\ \bibnamefont {Gould}}, \ and\ \bibinfo {author}
  {\bibfnamefont {H.~J.}\ \bibnamefont {Metcalf}},\ }\href {\doibase
  10.1103/PhysRevLett.61.169} {\bibfield  {journal} {\bibinfo  {journal} {Phys.
  Rev. Lett.}\ }\textbf {\bibinfo {volume} {61}},\ \bibinfo {pages} {169}
  (\bibinfo {year} {1988})}\BibitemShut {NoStop}%
\bibitem [{\citenamefont {Dalibard}\ and\ \citenamefont
  {Cohen-Tannoudji}(1989)}]{Dalibard1989Sysiphus}%
  \BibitemOpen
  \bibfield  {author} {\bibinfo {author} {\bibfnamefont {J.}~\bibnamefont
  {Dalibard}}\ and\ \bibinfo {author} {\bibfnamefont {C.}~\bibnamefont
  {Cohen-Tannoudji}},\ }\href
  {http://josab.osa.org/abstract.cfm?URI=josab-6-11-2023} {\bibfield  {journal}
  {\bibinfo  {journal} {J. Opt. Soc. Am. B}\ }\textbf {\bibinfo {volume} {6}},\
  \bibinfo {pages} {2023} (\bibinfo {year} {1989})}\BibitemShut {NoStop}%
\bibitem [{\citenamefont {Salomon}\ \emph {et~al.}(2013)\citenamefont
  {Salomon}, \citenamefont {Fouche}, \citenamefont {Wang}, \citenamefont
  {Aspect}, \citenamefont {Bouyer},\ and\ \citenamefont
  {Bourdel}}]{Salomon2013K39GrayMollasses}%
  \BibitemOpen
  \bibfield  {author} {\bibinfo {author} {\bibfnamefont {G.}~\bibnamefont
  {Salomon}}, \bibinfo {author} {\bibfnamefont {L.}~\bibnamefont {Fouche}},
  \bibinfo {author} {\bibfnamefont {P.}~\bibnamefont {Wang}}, \bibinfo {author}
  {\bibfnamefont {A.}~\bibnamefont {Aspect}}, \bibinfo {author} {\bibfnamefont
  {P.}~\bibnamefont {Bouyer}}, \ and\ \bibinfo {author} {\bibfnamefont
  {T.}~\bibnamefont {Bourdel}},\ }\href
  {http://stacks.iop.org/0295-5075/104/i=6/a=63002} {\bibfield  {journal}
  {\bibinfo  {journal} {Europhys. Lett.}\ }\textbf {\bibinfo {volume} {104}},\
  \bibinfo {pages} {63002} (\bibinfo {year} {2013})}\BibitemShut {NoStop}%
\bibitem [{\citenamefont {Colzi}\ \emph {et~al.}(2016)\citenamefont {Colzi},
  \citenamefont {Durastante}, \citenamefont {Fava}, \citenamefont {Serafini},
  \citenamefont {Lamporesi},\ and\ \citenamefont
  {Ferrari}}]{Colzi2016NaGrayMolasses}%
  \BibitemOpen
  \bibfield  {author} {\bibinfo {author} {\bibfnamefont {G.}~\bibnamefont
  {Colzi}}, \bibinfo {author} {\bibfnamefont {G.}~\bibnamefont {Durastante}},
  \bibinfo {author} {\bibfnamefont {E.}~\bibnamefont {Fava}}, \bibinfo {author}
  {\bibfnamefont {S.}~\bibnamefont {Serafini}}, \bibinfo {author}
  {\bibfnamefont {G.}~\bibnamefont {Lamporesi}}, \ and\ \bibinfo {author}
  {\bibfnamefont {G.}~\bibnamefont {Ferrari}},\ }\href {\doibase
  10.1103/PhysRevA.93.023421} {\bibfield  {journal} {\bibinfo  {journal} {Phys.
  Rev. A}\ }\textbf {\bibinfo {volume} {93}},\ \bibinfo {pages} {023421}
  (\bibinfo {year} {2016})}\BibitemShut {NoStop}%
\bibitem [{\citenamefont {Kasevich}\ and\ \citenamefont
  {Chu}(1992)}]{Kasevich1992RamanCooling}%
  \BibitemOpen
  \bibfield  {author} {\bibinfo {author} {\bibfnamefont {M.}~\bibnamefont
  {Kasevich}}\ and\ \bibinfo {author} {\bibfnamefont {S.}~\bibnamefont {Chu}},\
  }\href {\doibase 10.1103/PhysRevLett.69.1741} {\bibfield  {journal} {\bibinfo
   {journal} {Phys. Rev. Lett.}\ }\textbf {\bibinfo {volume} {69}},\ \bibinfo
  {pages} {1741} (\bibinfo {year} {1992})}\BibitemShut {NoStop}%
\bibitem [{\citenamefont {Aspect}\ \emph {et~al.}(1988)\citenamefont {Aspect},
  \citenamefont {Arimondo}, \citenamefont {Kaiser}, \citenamefont
  {Vansteenkiste},\ and\ \citenamefont
  {Cohen-Tannoudji}}]{Aspect1988VelocSelect}%
  \BibitemOpen
  \bibfield  {author} {\bibinfo {author} {\bibfnamefont {A.}~\bibnamefont
  {Aspect}}, \bibinfo {author} {\bibfnamefont {E.}~\bibnamefont {Arimondo}},
  \bibinfo {author} {\bibfnamefont {R.}~\bibnamefont {Kaiser}}, \bibinfo
  {author} {\bibfnamefont {N.}~\bibnamefont {Vansteenkiste}}, \ and\ \bibinfo
  {author} {\bibfnamefont {C.}~\bibnamefont {Cohen-Tannoudji}},\ }\href
  {\doibase 10.1103/PhysRevLett.61.826} {\bibfield  {journal} {\bibinfo
  {journal} {Phys. Rev. Lett.}\ }\textbf {\bibinfo {volume} {61}},\ \bibinfo
  {pages} {826} (\bibinfo {year} {1988})}\BibitemShut {NoStop}%
\bibitem [{\citenamefont {R\"{u}hrig}\ \emph {et~al.}(2015)\citenamefont
  {R\"{u}hrig}, \citenamefont {B\"{a}uerle}, \citenamefont {Griesmaier},\ and\
  \citenamefont {Pfau}}]{Ruhrig2015DemagnCoolingLimit}%
  \BibitemOpen
  \bibfield  {author} {\bibinfo {author} {\bibfnamefont {J.}~\bibnamefont
  {R\"{u}hrig}}, \bibinfo {author} {\bibfnamefont {T.}~\bibnamefont
  {B\"{a}uerle}}, \bibinfo {author} {\bibfnamefont {A.}~\bibnamefont
  {Griesmaier}}, \ and\ \bibinfo {author} {\bibfnamefont {T.}~\bibnamefont
  {Pfau}},\ }\href {\doibase 10.1364/OE.23.005596} {\bibfield  {journal}
  {\bibinfo  {journal} {Opt. Express}\ }\textbf {\bibinfo {volume} {23}},\
  \bibinfo {pages} {5596} (\bibinfo {year} {2015})}\BibitemShut {NoStop}%
\bibitem [{\citenamefont {Hamann}\ \emph {et~al.}(1998)\citenamefont {Hamann},
  \citenamefont {Haycock}, \citenamefont {Klose}, \citenamefont {Pax},
  \citenamefont {Deutsch},\ and\ \citenamefont
  {Jessen}}]{Hamann1998ResolvRamanSideband}%
  \BibitemOpen
  \bibfield  {author} {\bibinfo {author} {\bibfnamefont {S.~E.}\ \bibnamefont
  {Hamann}}, \bibinfo {author} {\bibfnamefont {D.~L.}\ \bibnamefont {Haycock}},
  \bibinfo {author} {\bibfnamefont {G.}~\bibnamefont {Klose}}, \bibinfo
  {author} {\bibfnamefont {P.~H.}\ \bibnamefont {Pax}}, \bibinfo {author}
  {\bibfnamefont {I.~H.}\ \bibnamefont {Deutsch}}, \ and\ \bibinfo {author}
  {\bibfnamefont {P.~S.}\ \bibnamefont {Jessen}},\ }\href
  {http://link.aps.org/doi/10.1103/PhysRevLett.80.4149} {\bibfield  {journal}
  {\bibinfo  {journal} {Phys. Rev. Lett.}\ }\textbf {\bibinfo {volume} {80}},\
  \bibinfo {pages} {4149} (\bibinfo {year} {1998})}\BibitemShut {NoStop}%
\bibitem [{\citenamefont {Kerman}\ \emph {et~al.}(2000)\citenamefont {Kerman},
  \citenamefont {Vuleti\ifmmode~\acute{c}\else \'{c}\fi{}}, \citenamefont
  {Chin},\ and\ \citenamefont {Chu}}]{Kerman2000RamanSideband3D}%
  \BibitemOpen
  \bibfield  {author} {\bibinfo {author} {\bibfnamefont {A.~J.}\ \bibnamefont
  {Kerman}}, \bibinfo {author} {\bibfnamefont {V.}~\bibnamefont
  {Vuleti\ifmmode~\acute{c}\else \'{c}\fi{}}}, \bibinfo {author} {\bibfnamefont
  {C.}~\bibnamefont {Chin}}, \ and\ \bibinfo {author} {\bibfnamefont
  {S.}~\bibnamefont {Chu}},\ }\href {\doibase 10.1103/PhysRevLett.84.439}
  {\bibfield  {journal} {\bibinfo  {journal} {Phys. Rev. Lett.}\ }\textbf
  {\bibinfo {volume} {84}},\ \bibinfo {pages} {439} (\bibinfo {year}
  {2000})}\BibitemShut {NoStop}%
\bibitem [{\citenamefont {Han}\ \emph {et~al.}(2000)\citenamefont {Han},
  \citenamefont {Wolf}, \citenamefont {Oliver}, \citenamefont {McCormick},
  \citenamefont {DePue},\ and\ \citenamefont
  {Weiss}}]{Han2000RamanSidebandCoolingWeiss}%
  \BibitemOpen
  \bibfield  {author} {\bibinfo {author} {\bibfnamefont {D.-J.}\ \bibnamefont
  {Han}}, \bibinfo {author} {\bibfnamefont {S.}~\bibnamefont {Wolf}}, \bibinfo
  {author} {\bibfnamefont {S.}~\bibnamefont {Oliver}}, \bibinfo {author}
  {\bibfnamefont {C.}~\bibnamefont {McCormick}}, \bibinfo {author}
  {\bibfnamefont {M.~T.}\ \bibnamefont {DePue}}, \ and\ \bibinfo {author}
  {\bibfnamefont {D.~S.}\ \bibnamefont {Weiss}},\ }\href {\doibase
  10.1103/PhysRevLett.85.724} {\bibfield  {journal} {\bibinfo  {journal} {Phys.
  Rev. Lett.}\ }\textbf {\bibinfo {volume} {85}},\ \bibinfo {pages} {724}
  (\bibinfo {year} {2000})}\BibitemShut {NoStop}%
\bibitem [{\citenamefont {Yang}\ \emph {et~al.}(2007)\citenamefont {Yang},
  \citenamefont {Halder}, \citenamefont {Appel}, \citenamefont {Hansen},\ and\
  \citenamefont {Hemmerich}}]{Yang2007ContinuousCaDTLoading}%
  \BibitemOpen
  \bibfield  {author} {\bibinfo {author} {\bibfnamefont {C.~Y.}\ \bibnamefont
  {Yang}}, \bibinfo {author} {\bibfnamefont {P.}~\bibnamefont {Halder}},
  \bibinfo {author} {\bibfnamefont {O.}~\bibnamefont {Appel}}, \bibinfo
  {author} {\bibfnamefont {D.}~\bibnamefont {Hansen}}, \ and\ \bibinfo {author}
  {\bibfnamefont {A.}~\bibnamefont {Hemmerich}},\ }\href {\doibase
  10.1103/PhysRevA.76.033418} {\bibfield  {journal} {\bibinfo  {journal} {Phys.
  Rev. A}\ }\textbf {\bibinfo {volume} {76}},\ \bibinfo {pages} {033418}
  (\bibinfo {year} {2007})}\BibitemShut {NoStop}%
\bibitem [{\citenamefont {Radwell}\ \emph {et~al.}(2013)\citenamefont
  {Radwell}, \citenamefont {Walker},\ and\ \citenamefont
  {Franke-Arnold}}]{Radwell2013HighPSDDarkSPOT}%
  \BibitemOpen
  \bibfield  {author} {\bibinfo {author} {\bibfnamefont {N.}~\bibnamefont
  {Radwell}}, \bibinfo {author} {\bibfnamefont {G.}~\bibnamefont {Walker}}, \
  and\ \bibinfo {author} {\bibfnamefont {S.}~\bibnamefont {Franke-Arnold}},\
  }\href {\doibase 10.1103/PhysRevA.88.043409} {\bibfield  {journal} {\bibinfo
  {journal} {Phys. Rev. A}\ }\textbf {\bibinfo {volume} {88}},\ \bibinfo
  {pages} {043409} (\bibinfo {year} {2013})}\BibitemShut {NoStop}%
\bibitem [{\citenamefont {Katori}\ \emph {et~al.}(1999)\citenamefont {Katori},
  \citenamefont {Ido}, \citenamefont {Isoya},\ and\ \citenamefont
  {Kuwata-Gonokami}}]{Katori1999SrMOTRecoilTemp}%
  \BibitemOpen
  \bibfield  {author} {\bibinfo {author} {\bibfnamefont {H.}~\bibnamefont
  {Katori}}, \bibinfo {author} {\bibfnamefont {T.}~\bibnamefont {Ido}},
  \bibinfo {author} {\bibfnamefont {Y.}~\bibnamefont {Isoya}}, \ and\ \bibinfo
  {author} {\bibfnamefont {M.}~\bibnamefont {Kuwata-Gonokami}},\ }\href
  {\doibase 10.1103/PhysRevLett.82.1116} {\bibfield  {journal} {\bibinfo
  {journal} {Phys. Rev. Lett.}\ }\textbf {\bibinfo {volume} {82}},\ \bibinfo
  {pages} {1116} (\bibinfo {year} {1999})}\BibitemShut {NoStop}%
\bibitem [{\citenamefont {Hu}\ \emph {et~al.}(2017)\citenamefont {Hu},
  \citenamefont {Urvoy}, \citenamefont {Vendeiro}, \citenamefont {Cr{\'e}pel},
  \citenamefont {Chen},\ and\ \citenamefont
  {Vuleti{\'c}}}]{Hu2013LaserCoolingBEC}%
  \BibitemOpen
  \bibfield  {author} {\bibinfo {author} {\bibfnamefont {J.}~\bibnamefont
  {Hu}}, \bibinfo {author} {\bibfnamefont {A.}~\bibnamefont {Urvoy}}, \bibinfo
  {author} {\bibfnamefont {Z.}~\bibnamefont {Vendeiro}}, \bibinfo {author}
  {\bibfnamefont {V.}~\bibnamefont {Cr{\'e}pel}}, \bibinfo {author}
  {\bibfnamefont {W.}~\bibnamefont {Chen}}, \ and\ \bibinfo {author}
  {\bibfnamefont {V.}~\bibnamefont {Vuleti{\'c}}},\ }\href {\doibase
  10.1126/science.aan5614} {\bibfield  {journal} {\bibinfo  {journal}
  {Science}\ }\textbf {\bibinfo {volume} {358}},\ \bibinfo {pages} {1078}
  (\bibinfo {year} {2017})},\ \Eprint
  {http://arxiv.org/abs/http://science.sciencemag.org/content/358/6366/1078.full.pdf}
  {http://science.sciencemag.org/content/358/6366/1078.full.pdf} \BibitemShut
  {NoStop}%
\bibitem [{\citenamefont {Kawasaki}\ \emph {et~al.}(2015)\citenamefont
  {Kawasaki}, \citenamefont {Braverman}, \citenamefont {Yu},\ and\
  \citenamefont {Vuleti\ifmmode~\acute{c}\else
  \'{c}\fi{}}}]{Kawasaki2015TwoColorYbMOT}%
  \BibitemOpen
  \bibfield  {author} {\bibinfo {author} {\bibfnamefont {A.}~\bibnamefont
  {Kawasaki}}, \bibinfo {author} {\bibfnamefont {B.}~\bibnamefont {Braverman}},
  \bibinfo {author} {\bibfnamefont {Q.}~\bibnamefont {Yu}}, \ and\ \bibinfo
  {author} {\bibfnamefont {V.}~\bibnamefont {Vuleti\ifmmode~\acute{c}\else
  \'{c}\fi{}}},\ }\href {http://stacks.iop.org/0953-4075/48/i=15/a=155302}
  {\bibfield  {journal} {\bibinfo  {journal} {J. Phys. B}\ }\textbf {\bibinfo
  {volume} {48}},\ \bibinfo {pages} {155302} (\bibinfo {year}
  {2015})}\BibitemShut {NoStop}%
\bibitem [{\citenamefont {Lee}\ \emph {et~al.}(2015)\citenamefont {Lee},
  \citenamefont {Lee}, \citenamefont {Noh},\ and\ \citenamefont
  {Mun}}]{Lee2015CoreShellYbMOT}%
  \BibitemOpen
  \bibfield  {author} {\bibinfo {author} {\bibfnamefont {J.}~\bibnamefont
  {Lee}}, \bibinfo {author} {\bibfnamefont {J.~H.}\ \bibnamefont {Lee}},
  \bibinfo {author} {\bibfnamefont {J.}~\bibnamefont {Noh}}, \ and\ \bibinfo
  {author} {\bibfnamefont {J.}~\bibnamefont {Mun}},\ }\href {\doibase
  10.1103/PhysRevA.91.053405} {\bibfield  {journal} {\bibinfo  {journal} {Phys.
  Rev. A}\ }\textbf {\bibinfo {volume} {91}},\ \bibinfo {pages} {053405}
  (\bibinfo {year} {2015})}\BibitemShut {NoStop}%
\bibitem [{\citenamefont {Gr\"unert}\ and\ \citenamefont
  {Hemmerich}(2002)}]{Hemmerich2002CaMOT}%
  \BibitemOpen
  \bibfield  {author} {\bibinfo {author} {\bibfnamefont {J.}~\bibnamefont
  {Gr\"unert}}\ and\ \bibinfo {author} {\bibfnamefont {A.}~\bibnamefont
  {Hemmerich}},\ }\href {\doibase 10.1103/PhysRevA.65.041401} {\bibfield
  {journal} {\bibinfo  {journal} {Phys. Rev. A}\ }\textbf {\bibinfo {volume}
  {65}},\ \bibinfo {pages} {041401} (\bibinfo {year} {2002})}\BibitemShut
  {NoStop}%
\bibitem [{\citenamefont {D\"orscher}\ \emph {et~al.}(2013)\citenamefont
  {D\"orscher}, \citenamefont {Thobe}, \citenamefont {Hundt}, \citenamefont
  {Kochanke}, \citenamefont {Targat}, \citenamefont {Windpassinger},
  \citenamefont {Becker},\ and\ \citenamefont
  {Sengstock}}]{Dorscher2013Yb2D3DMOT}%
  \BibitemOpen
  \bibfield  {author} {\bibinfo {author} {\bibfnamefont {S.}~\bibnamefont
  {D\"orscher}}, \bibinfo {author} {\bibfnamefont {A.}~\bibnamefont {Thobe}},
  \bibinfo {author} {\bibfnamefont {B.}~\bibnamefont {Hundt}}, \bibinfo
  {author} {\bibfnamefont {A.}~\bibnamefont {Kochanke}}, \bibinfo {author}
  {\bibfnamefont {R.~L.}\ \bibnamefont {Targat}}, \bibinfo {author}
  {\bibfnamefont {P.}~\bibnamefont {Windpassinger}}, \bibinfo {author}
  {\bibfnamefont {C.}~\bibnamefont {Becker}}, \ and\ \bibinfo {author}
  {\bibfnamefont {K.}~\bibnamefont {Sengstock}},\ }\href {\doibase
  10.1063/1.4802682} {\bibfield  {journal} {\bibinfo  {journal} {Rev. Sci.
  Instrum.}\ }\textbf {\bibinfo {volume} {84}},\ \bibinfo {pages} {043109}
  (\bibinfo {year} {2013})}\BibitemShut {NoStop}%
\bibitem [{\citenamefont {Frisch}\ \emph {et~al.}(2012)\citenamefont {Frisch},
  \citenamefont {Aikawa}, \citenamefont {Mark}, \citenamefont {Rietzler},
  \citenamefont {Schindler}, \citenamefont {Zupani\ifmmode~\check{c}\else
  \v{c}\fi{}}, \citenamefont {Grimm},\ and\ \citenamefont
  {Ferlaino}}]{Frisch2012ErNarrowLineMOT}%
  \BibitemOpen
  \bibfield  {author} {\bibinfo {author} {\bibfnamefont {A.}~\bibnamefont
  {Frisch}}, \bibinfo {author} {\bibfnamefont {K.}~\bibnamefont {Aikawa}},
  \bibinfo {author} {\bibfnamefont {M.}~\bibnamefont {Mark}}, \bibinfo {author}
  {\bibfnamefont {A.}~\bibnamefont {Rietzler}}, \bibinfo {author}
  {\bibfnamefont {J.}~\bibnamefont {Schindler}}, \bibinfo {author}
  {\bibfnamefont {E.}~\bibnamefont {Zupani\ifmmode~\check{c}\else \v{c}\fi{}}},
  \bibinfo {author} {\bibfnamefont {R.}~\bibnamefont {Grimm}}, \ and\ \bibinfo
  {author} {\bibfnamefont {F.}~\bibnamefont {Ferlaino}},\ }\href {\doibase
  10.1103/PhysRevA.85.051401} {\bibfield  {journal} {\bibinfo  {journal} {Phys.
  Rev. A}\ }\textbf {\bibinfo {volume} {85}},\ \bibinfo {pages} {051401}
  (\bibinfo {year} {2012})}\BibitemShut {NoStop}%
\bibitem [{\citenamefont {Ketterle}\ \emph {et~al.}(1993)\citenamefont
  {Ketterle}, \citenamefont {Davis}, \citenamefont {Joffe}, \citenamefont
  {Martin},\ and\ \citenamefont {Pritchard}}]{Ketterle1993FirstDarkSPOT}%
  \BibitemOpen
  \bibfield  {author} {\bibinfo {author} {\bibfnamefont {W.}~\bibnamefont
  {Ketterle}}, \bibinfo {author} {\bibfnamefont {K.~B.}\ \bibnamefont {Davis}},
  \bibinfo {author} {\bibfnamefont {M.~A.}\ \bibnamefont {Joffe}}, \bibinfo
  {author} {\bibfnamefont {A.}~\bibnamefont {Martin}}, \ and\ \bibinfo {author}
  {\bibfnamefont {D.~E.}\ \bibnamefont {Pritchard}},\ }\href {\doibase
  10.1103/PhysRevLett.70.2253} {\bibfield  {journal} {\bibinfo  {journal}
  {Phys. Rev. Lett.}\ }\textbf {\bibinfo {volume} {70}},\ \bibinfo {pages}
  {2253} (\bibinfo {year} {1993})}\BibitemShut {NoStop}%
\bibitem [{\citenamefont {Mahnke}\ \emph {et~al.}(2015)\citenamefont {Mahnke},
  \citenamefont {Kruse}, \citenamefont {H\"uper}, \citenamefont {J\"ollenbeck},
  \citenamefont {Ertmer}, \citenamefont {Arlt},\ and\ \citenamefont
  {Klempt}}]{Mahnke2015ContinuousMagnReservoir}%
  \BibitemOpen
  \bibfield  {author} {\bibinfo {author} {\bibfnamefont {J.}~\bibnamefont
  {Mahnke}}, \bibinfo {author} {\bibfnamefont {I.}~\bibnamefont {Kruse}},
  \bibinfo {author} {\bibfnamefont {A.}~\bibnamefont {H\"uper}}, \bibinfo
  {author} {\bibfnamefont {S.}~\bibnamefont {J\"ollenbeck}}, \bibinfo {author}
  {\bibfnamefont {W.}~\bibnamefont {Ertmer}}, \bibinfo {author} {\bibfnamefont
  {J.}~\bibnamefont {Arlt}}, \ and\ \bibinfo {author} {\bibfnamefont
  {C.}~\bibnamefont {Klempt}},\ }\href
  {http://stacks.iop.org/0953-4075/48/i=16/a=165301} {\bibfield  {journal}
  {\bibinfo  {journal} {J. Phys. B}\ }\textbf {\bibinfo {volume} {48}},\
  \bibinfo {pages} {165301} (\bibinfo {year} {2015})}\BibitemShut {NoStop}%
\bibitem [{\citenamefont {Falkenau}\ \emph {et~al.}(2011)\citenamefont
  {Falkenau}, \citenamefont {Volchkov}, \citenamefont {R\"uhrig}, \citenamefont
  {Griesmaier},\ and\ \citenamefont {Pfau}}]{Falkenau2011ContinuousDTLoading}%
  \BibitemOpen
  \bibfield  {author} {\bibinfo {author} {\bibfnamefont {M.}~\bibnamefont
  {Falkenau}}, \bibinfo {author} {\bibfnamefont {V.~V.}\ \bibnamefont
  {Volchkov}}, \bibinfo {author} {\bibfnamefont {J.}~\bibnamefont {R\"uhrig}},
  \bibinfo {author} {\bibfnamefont {A.}~\bibnamefont {Griesmaier}}, \ and\
  \bibinfo {author} {\bibfnamefont {T.}~\bibnamefont {Pfau}},\ }\href {\doibase
  10.1103/PhysRevLett.106.163002} {\bibfield  {journal} {\bibinfo  {journal}
  {Phys. Rev. Lett.}\ }\textbf {\bibinfo {volume} {106}},\ \bibinfo {pages}
  {163002} (\bibinfo {year} {2011})}\BibitemShut {NoStop}%
\bibitem [{\citenamefont {Volchkov}\ \emph {et~al.}(2013)\citenamefont
  {Volchkov}, \citenamefont {R\''uhrig}, \citenamefont {Pfau},\ and\
  \citenamefont {Griesmaier}}]{Volchkov2013sci}%
  \BibitemOpen
  \bibfield  {author} {\bibinfo {author} {\bibfnamefont {V.~V.}\ \bibnamefont
  {Volchkov}}, \bibinfo {author} {\bibfnamefont {J.}~\bibnamefont {R\''uhrig}},
  \bibinfo {author} {\bibfnamefont {T.}~\bibnamefont {Pfau}}, \ and\ \bibinfo
  {author} {\bibfnamefont {A.}~\bibnamefont {Griesmaier}},\ }\href
  {http://stacks.iop.org/1367-2630/15/i=9/a=093012} {\bibfield  {journal}
  {\bibinfo  {journal} {New J. Phys.}\ }\textbf {\bibinfo {volume} {15}},\
  \bibinfo {pages} {093012} (\bibinfo {year} {2013})}\BibitemShut {NoStop}%
\bibitem [{\citenamefont {Chikkatur}\ \emph {et~al.}(2002)\citenamefont
  {Chikkatur}, \citenamefont {Shin}, \citenamefont {Leanhardt}, \citenamefont
  {Kielpinski}, \citenamefont {Tsikata}, \citenamefont {Gustavson},
  \citenamefont {Pritchard},\ and\ \citenamefont
  {Ketterle}}]{Chikkatur2002ContinuousBEC}%
  \BibitemOpen
  \bibfield  {author} {\bibinfo {author} {\bibfnamefont {A.~P.}\ \bibnamefont
  {Chikkatur}}, \bibinfo {author} {\bibfnamefont {Y.}~\bibnamefont {Shin}},
  \bibinfo {author} {\bibfnamefont {A.~E.}\ \bibnamefont {Leanhardt}}, \bibinfo
  {author} {\bibfnamefont {D.}~\bibnamefont {Kielpinski}}, \bibinfo {author}
  {\bibfnamefont {E.}~\bibnamefont {Tsikata}}, \bibinfo {author} {\bibfnamefont
  {T.~L.}\ \bibnamefont {Gustavson}}, \bibinfo {author} {\bibfnamefont {D.~E.}\
  \bibnamefont {Pritchard}}, \ and\ \bibinfo {author} {\bibfnamefont
  {W.}~\bibnamefont {Ketterle}},\ }\href
  {http://www.sciencemag.org/content/296/5576/2193.abstract} {\bibfield
  {journal} {\bibinfo  {journal} {Science}\ }\textbf {\bibinfo {volume}
  {296}},\ \bibinfo {pages} {2193} (\bibinfo {year} {2002})}\BibitemShut
  {NoStop}%
\bibitem [{\citenamefont {Robins}\ \emph {et~al.}(2008)\citenamefont {Robins},
  \citenamefont {Figl}, \citenamefont {Jeppesen}, \citenamefont {Dennis},\ and\
  \citenamefont {Close}}]{Robins2008PumpedAtomLaser}%
  \BibitemOpen
  \bibfield  {author} {\bibinfo {author} {\bibfnamefont {N.~P.}\ \bibnamefont
  {Robins}}, \bibinfo {author} {\bibfnamefont {C.}~\bibnamefont {Figl}},
  \bibinfo {author} {\bibfnamefont {M.}~\bibnamefont {Jeppesen}}, \bibinfo
  {author} {\bibfnamefont {G.~R.}\ \bibnamefont {Dennis}}, \ and\ \bibinfo
  {author} {\bibfnamefont {J.~D.}\ \bibnamefont {Close}},\ }\href {\doibase
  10.1038/nphys1027} {\bibfield  {journal} {\bibinfo  {journal} {Nat. Phys.}\
  }\textbf {\bibinfo {volume} {4}},\ \bibinfo {pages} {731} (\bibinfo {year}
  {2008})}\BibitemShut {NoStop}%
\bibitem [{\citenamefont {Stellmer}(2013)}]{StellmerPhD}%
  \BibitemOpen
  \bibfield  {author} {\bibinfo {author} {\bibfnamefont {S.}~\bibnamefont
  {Stellmer}},\ }\href@noop {} {Ph.D. thesis},\ \bibinfo  {school} {University
  of Innsbruck} (\bibinfo {year} {2013})\BibitemShut {NoStop}%
\bibitem [{\citenamefont {Stellmer}\ \emph {et~al.}(2009)\citenamefont
  {Stellmer}, \citenamefont {Tey}, \citenamefont {Huang}, \citenamefont
  {Grimm},\ and\ \citenamefont {Schreck}}]{Stellmer2009bec}%
  \BibitemOpen
  \bibfield  {author} {\bibinfo {author} {\bibfnamefont {S.}~\bibnamefont
  {Stellmer}}, \bibinfo {author} {\bibfnamefont {M.~K.}\ \bibnamefont {Tey}},
  \bibinfo {author} {\bibfnamefont {B.}~\bibnamefont {Huang}}, \bibinfo
  {author} {\bibfnamefont {R.}~\bibnamefont {Grimm}}, \ and\ \bibinfo {author}
  {\bibfnamefont {F.}~\bibnamefont {Schreck}},\ }\href {\doibase
  10.1103/PhysRevLett.103.200401} {\bibfield  {journal} {\bibinfo  {journal}
  {Phys. Rev. Lett.}\ }\textbf {\bibinfo {volume} {103}},\ \bibinfo {pages}
  {200401} (\bibinfo {year} {2009})}\BibitemShut {NoStop}%
\bibitem [{Sup()}]{SupplInf}%
  \BibitemOpen
  \href@noop {} {\bibinfo  {journal} {See Supplemental Material}\ }\BibitemShut
  {NoStop}%
\bibitem [{\citenamefont {Senaratne}\ \emph {et~al.}(2015)\citenamefont
  {Senaratne}, \citenamefont {Rajagopal}, \citenamefont {Geiger}, \citenamefont
  {Fujiwara}, \citenamefont {Lebedev},\ and\ \citenamefont
  {Weld}}]{Senaratne2015OvenMicrotubes}%
  \BibitemOpen
\bibfield  {journal} {  }\bibfield  {author} {\bibinfo {author} {\bibfnamefont
  {R.}~\bibnamefont {Senaratne}}, \bibinfo {author} {\bibfnamefont {S.~V.}\
  \bibnamefont {Rajagopal}}, \bibinfo {author} {\bibfnamefont {Z.~A.}\
  \bibnamefont {Geiger}}, \bibinfo {author} {\bibfnamefont {K.~M.}\
  \bibnamefont {Fujiwara}}, \bibinfo {author} {\bibfnamefont {V.}~\bibnamefont
  {Lebedev}}, \ and\ \bibinfo {author} {\bibfnamefont {D.~M.}\ \bibnamefont
  {Weld}},\ }\href {\doibase 10.1063/1.4907401} {\bibfield  {journal} {\bibinfo
   {journal} {Rev. Sci. Instrum.}\ }\textbf {\bibinfo {volume} {86}},\ \bibinfo
  {pages} {023105} (\bibinfo {year} {2015})}\BibitemShut {NoStop}%
\bibitem [{\citenamefont {Dieckmann}\ \emph {et~al.}(1998)\citenamefont
  {Dieckmann}, \citenamefont {Spreeuw}, \citenamefont {Weidem\"uller},\ and\
  \citenamefont {Walraven}}]{Dieckmann19982DMOT}%
  \BibitemOpen
  \bibfield  {author} {\bibinfo {author} {\bibfnamefont {K.}~\bibnamefont
  {Dieckmann}}, \bibinfo {author} {\bibfnamefont {R.~J.~C.}\ \bibnamefont
  {Spreeuw}}, \bibinfo {author} {\bibfnamefont {M.}~\bibnamefont
  {Weidem\"uller}}, \ and\ \bibinfo {author} {\bibfnamefont {J.~T.~M.}\
  \bibnamefont {Walraven}},\ }\href {\doibase 10.1103/PhysRevA.58.3891}
  {\bibfield  {journal} {\bibinfo  {journal} {Phys. Rev. A}\ }\textbf {\bibinfo
  {volume} {58}},\ \bibinfo {pages} {3891} (\bibinfo {year}
  {1998})}\BibitemShut {NoStop}%
\bibitem [{\citenamefont {Yang}\ \emph {et~al.}(2015)\citenamefont {Yang},
  \citenamefont {Pandey}, \citenamefont {Pramod}, \citenamefont {Leroux},
  \citenamefont {Kwong}, \citenamefont {Hajiyev}, \citenamefont {Chia},
  \citenamefont {Fang},\ and\ \citenamefont
  {Wilkowski}}]{Wilkowski20152DSrMOT}%
  \BibitemOpen
  \bibfield  {author} {\bibinfo {author} {\bibfnamefont {T.}~\bibnamefont
  {Yang}}, \bibinfo {author} {\bibfnamefont {K.}~\bibnamefont {Pandey}},
  \bibinfo {author} {\bibfnamefont {M.~S.}\ \bibnamefont {Pramod}}, \bibinfo
  {author} {\bibfnamefont {F.}~\bibnamefont {Leroux}}, \bibinfo {author}
  {\bibfnamefont {C.~C.}\ \bibnamefont {Kwong}}, \bibinfo {author}
  {\bibfnamefont {E.}~\bibnamefont {Hajiyev}}, \bibinfo {author} {\bibfnamefont
  {Z.~Y.}\ \bibnamefont {Chia}}, \bibinfo {author} {\bibfnamefont
  {B.}~\bibnamefont {Fang}}, \ and\ \bibinfo {author} {\bibfnamefont
  {D.}~\bibnamefont {Wilkowski}},\ }\href {\doibase 10.1140/epjd/e2015-60288-y}
  {\bibfield  {journal} {\bibinfo  {journal} {The Eur. Phys. J. D}\ }\textbf
  {\bibinfo {volume} {69}},\ \bibinfo {pages} {226} (\bibinfo {year}
  {2015})}\BibitemShut {NoStop}%
\bibitem [{\citenamefont {Nosske}\ \emph {et~al.}(2017)\citenamefont {Nosske},
  \citenamefont {Couturier}, \citenamefont {Hu}, \citenamefont {Tan},
  \citenamefont {Qiao}, \citenamefont {Blume}, \citenamefont {Jiang},
  \citenamefont {Chen},\ and\ \citenamefont
  {Weidem\"uller}}]{Weidemuller20172DSrMOT}%
  \BibitemOpen
  \bibfield  {author} {\bibinfo {author} {\bibfnamefont {I.}~\bibnamefont
  {Nosske}}, \bibinfo {author} {\bibfnamefont {L.}~\bibnamefont {Couturier}},
  \bibinfo {author} {\bibfnamefont {F.}~\bibnamefont {Hu}}, \bibinfo {author}
  {\bibfnamefont {C.}~\bibnamefont {Tan}}, \bibinfo {author} {\bibfnamefont
  {C.}~\bibnamefont {Qiao}}, \bibinfo {author} {\bibfnamefont {J.}~\bibnamefont
  {Blume}}, \bibinfo {author} {\bibfnamefont {Y.~H.}\ \bibnamefont {Jiang}},
  \bibinfo {author} {\bibfnamefont {P.}~\bibnamefont {Chen}}, \ and\ \bibinfo
  {author} {\bibfnamefont {M.}~\bibnamefont {Weidem\"uller}},\ }\href {\doibase
  10.1103/PhysRevA.96.053415} {\bibfield  {journal} {\bibinfo  {journal} {Phys.
  Rev. A}\ }\textbf {\bibinfo {volume} {96}},\ \bibinfo {pages} {053415}
  (\bibinfo {year} {2017})}\BibitemShut {NoStop}%
\bibitem [{\citenamefont {Tey}\ \emph {et~al.}(2010)\citenamefont {Tey},
  \citenamefont {Stellmer}, \citenamefont {Grimm},\ and\ \citenamefont
  {Schreck}}]{Tey2010DoubleBECFermiSea}%
  \BibitemOpen
  \bibfield  {author} {\bibinfo {author} {\bibfnamefont {M.~K.}\ \bibnamefont
  {Tey}}, \bibinfo {author} {\bibfnamefont {S.}~\bibnamefont {Stellmer}},
  \bibinfo {author} {\bibfnamefont {R.}~\bibnamefont {Grimm}}, \ and\ \bibinfo
  {author} {\bibfnamefont {F.}~\bibnamefont {Schreck}},\ }\href {\doibase
  10.1103/PhysRevA.82.011608} {\bibfield  {journal} {\bibinfo  {journal} {Phys.
  Rev. A}\ }\textbf {\bibinfo {volume} {82}},\ \bibinfo {pages} {011608}
  (\bibinfo {year} {2010})}\BibitemShut {NoStop}%
\bibitem [{\citenamefont {Prodan}\ \emph {et~al.}(1982)\citenamefont {Prodan},
  \citenamefont {Phillips},\ and\ \citenamefont
  {Metcalf}}]{Prodan1982MonoEnergBeam}%
  \BibitemOpen
  \bibfield  {author} {\bibinfo {author} {\bibfnamefont {J.~V.}\ \bibnamefont
  {Prodan}}, \bibinfo {author} {\bibfnamefont {W.~D.}\ \bibnamefont
  {Phillips}}, \ and\ \bibinfo {author} {\bibfnamefont {H.}~\bibnamefont
  {Metcalf}},\ }\href {https://link.aps.org/doi/10.1103/PhysRevLett.49.1149}
  {\bibfield  {journal} {\bibinfo  {journal} {Phys. Rev. Lett.}\ }\textbf
  {\bibinfo {volume} {49}},\ \bibinfo {pages} {1149} (\bibinfo {year}
  {1982})}\BibitemShut {NoStop}%
\bibitem [{\citenamefont {Zhu}\ \emph {et~al.}(1991)\citenamefont {Zhu},
  \citenamefont {Oates},\ and\ \citenamefont
  {Hall}}]{Zhu1991FirstExpWhiteLightSlowing}%
  \BibitemOpen
  \bibfield  {author} {\bibinfo {author} {\bibfnamefont {M.}~\bibnamefont
  {Zhu}}, \bibinfo {author} {\bibfnamefont {C.~W.}\ \bibnamefont {Oates}}, \
  and\ \bibinfo {author} {\bibfnamefont {J.~L.}\ \bibnamefont {Hall}},\ }\href
  {https://link.aps.org/doi/10.1103/PhysRevLett.67.46} {\bibfield  {journal}
  {\bibinfo  {journal} {Phys. Rev. Lett.}\ }\textbf {\bibinfo {volume} {67}},\
  \bibinfo {pages} {46} (\bibinfo {year} {1991})}\BibitemShut {NoStop}%
\bibitem [{\citenamefont {Borkowski}\ \emph {et~al.}(2014)\citenamefont
  {Borkowski}, \citenamefont {Morzy\ifmmode~\acute{n}\else \'{n}\fi{}ski},
  \citenamefont {Ciury\l{}o}, \citenamefont {Julienne}, \citenamefont {Yan},
  \citenamefont {DeSalvo},\ and\ \citenamefont
  {Killian}}]{Borkowski2014SrMassScalingPot}%
  \BibitemOpen
  \bibfield  {author} {\bibinfo {author} {\bibfnamefont {M.}~\bibnamefont
  {Borkowski}}, \bibinfo {author} {\bibfnamefont {P.}~\bibnamefont
  {Morzy\ifmmode~\acute{n}\else \'{n}\fi{}ski}}, \bibinfo {author}
  {\bibfnamefont {R.}~\bibnamefont {Ciury\l{}o}}, \bibinfo {author}
  {\bibfnamefont {P.~S.}\ \bibnamefont {Julienne}}, \bibinfo {author}
  {\bibfnamefont {M.}~\bibnamefont {Yan}}, \bibinfo {author} {\bibfnamefont
  {B.~J.}\ \bibnamefont {DeSalvo}}, \ and\ \bibinfo {author} {\bibfnamefont
  {T.~C.}\ \bibnamefont {Killian}},\ }\href {\doibase
  10.1103/PhysRevA.90.032713} {\bibfield  {journal} {\bibinfo  {journal} {Phys.
  Rev. A}\ }\textbf {\bibinfo {volume} {90}},\ \bibinfo {pages} {032713}
  (\bibinfo {year} {2014})}\BibitemShut {NoStop}%
\bibitem [{\citenamefont {Berglund}\ \emph {et~al.}(2008)\citenamefont
  {Berglund}, \citenamefont {Hanssen},\ and\ \citenamefont
  {McClelland}}]{Berglund2008NarrowLineErbiumMOT}%
  \BibitemOpen
  \bibfield  {author} {\bibinfo {author} {\bibfnamefont {A.~J.}\ \bibnamefont
  {Berglund}}, \bibinfo {author} {\bibfnamefont {J.~L.}\ \bibnamefont
  {Hanssen}}, \ and\ \bibinfo {author} {\bibfnamefont {J.~J.}\ \bibnamefont
  {McClelland}},\ }\href {\doibase 10.1103/PhysRevLett.100.113002} {\bibfield
  {journal} {\bibinfo  {journal} {Phys. Rev. Lett.}\ }\textbf {\bibinfo
  {volume} {100}},\ \bibinfo {pages} {113002} (\bibinfo {year}
  {2008})}\BibitemShut {NoStop}%
\bibitem [{\citenamefont {Kuwamoto}\ \emph {et~al.}(1999)\citenamefont
  {Kuwamoto}, \citenamefont {Honda}, \citenamefont {Takahashi},\ and\
  \citenamefont {Yabuzaki}}]{Kuwamoto1999NarrowLineMOTYb}%
  \BibitemOpen
  \bibfield  {author} {\bibinfo {author} {\bibfnamefont {T.}~\bibnamefont
  {Kuwamoto}}, \bibinfo {author} {\bibfnamefont {K.}~\bibnamefont {Honda}},
  \bibinfo {author} {\bibfnamefont {Y.}~\bibnamefont {Takahashi}}, \ and\
  \bibinfo {author} {\bibfnamefont {T.}~\bibnamefont {Yabuzaki}},\ }\href
  {\doibase 10.1103/PhysRevA.60.R745} {\bibfield  {journal} {\bibinfo
  {journal} {Phys. Rev. A}\ }\textbf {\bibinfo {volume} {60}},\ \bibinfo
  {pages} {R745} (\bibinfo {year} {1999})}\BibitemShut {NoStop}%
\end{thebibliography}

%

\setcounter{table}{0}
\renewcommand{\thetable}{S-\Roman{table}}
\setcounter{figure}{0}
\renewcommand{\thefigure}{S-\arabic{figure}}
\renewcommand{\bottomfraction}{0.85}
\renewcommand{\textfraction}{0.15}

\section{Supplemental Material}
\section{Details of the experimental setup}
\label{sec:Experiment}

The atomic source is an oven producing a $\unit[10]{mm}$ diameter effusive jet of strontium atoms by heating Sr metal pieces ($\unit[99.5]{\%}$ from Alfa Aesar) to around $\unit[500]{^\circ C}$. While the oven was designed to operate at $\unit[550]{^\circ C}$ for 18 months it is currently operated at only $\unit[500]{^\circ C}$ as this has been found to deliver sufficient flux and will extend the operating life before needing to refill the Sr reservoir and rebake the oven section of the vacuum system. Increasing the temperature to $\unit[550]{^\circ C}$ is expected to increase the flux by a factor of five, an important option when working with $\unit[0.5]{\%}$ abundant $^{84}\mathrm{Sr}$. The oven output is collimated by an array of $\unit[8]{mm}$ long stainless steel microtubes with $\unit[80]{\mu m}$ inner diameter and $\unit[180]{\mu m}$ outer diameter, following a design inspired by \cite{Senaratne2015OvenMicrotubes}. The atoms are then transversely cooled in two dimensions over a $\unit[90]{mm}$ long region using for each axis a ``zig-zag" configuration with four passes and retro-reflection of the laser beam. Finally a $\unit[105]{cm}$ long Zeeman slower, with a spin-flip design and a modeled maximum capture velocity of $\unit[500]{m/s}$, slows the atoms to approximately $\unit[20]{m/s}$.

\begin{figure*}[tb]
\centering
\includegraphics[width=0.80\textwidth]{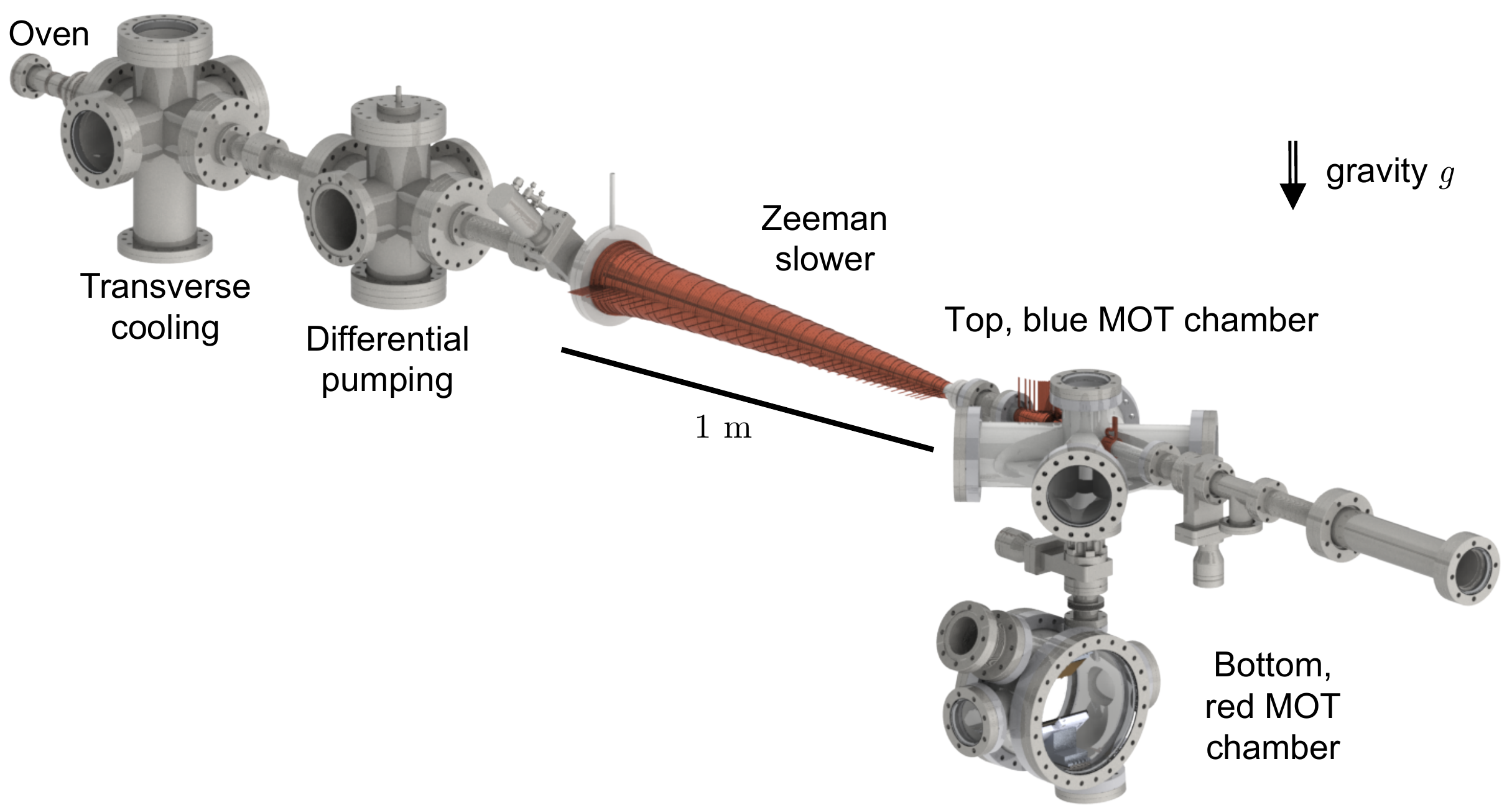}
\caption{\label{fig:SM1_MachineCADDiagram} CAD drawing of the machine excluding vacuum pumping sections.}
\end{figure*}

The Zeeman slower output is captured by a 2D MOT using the blue $\unit[461]{nm}$ transition. The 2D MOT magnetic field gradient is created by two vertical permanent magnet arrays, facing each other with opposite pole orientation. The position and orientation of individual magnets stacked in each array can be independently adjusted to produce the desired radial gradient, which is approximately $\unit[10]{G / cm}$ in the region of the blue MOT and rapidly transitions to approximately $\unit[2]{G / cm}$ over about $\unit[15]{mm}$. Helmholtz bias coils are used to trim the pointing and location of the 2D MOT in order to maximize capture by the 3D red MOT in the second chamber. Other details are described in the main text and illustrated in Figure~\ref{fig:SM1_MachineCADDiagram}. The parameters of all the beams for laser cooling on the blue transition are listed in Table~\ref{tab:LaserParameters0}. The parameters of the beams used for Red MOT I, Red MOT II(a) and Red MOT II(b) are listed in Table~\ref{tab:LaserParameters1}, Table~\ref{tab:LaserParameters2} and Table~\ref{tab:LaserParameters3}, respectively.

\begin{table*}
\caption{Properties of laser cooling beams using the blue ${^1\mathrm{S}_0} - {^1\mathrm{P}_1}$ transition.}
\label{tab:LaserParameters0}
\begin{ruledtabular}
\begin{tabular}{C{0.13\textwidth}C{0.15\textwidth}C{0.1\textwidth}C{0.1\textwidth}C{0.3\textwidth}}
\textbf{Beam name} & \textbf{Detuning [MHz]} & \textbf{Power [mW]} & \textbf{$1/e^2$ diameter [mm]} & \textbf{Comments}\\
\noalign{\smallskip}\hline\hline\noalign{\smallskip}
Transverse cooling & $-18$ ($-0.6 \Gamma$) & 30 & $23 \times 9.6$ & Beam is passed 4 times side by side giving a $\unit[90]{mm}$ long cooling region. Same for both horizontal and vertical axes.\\
\hline
Zeeman slower & $-424$ ($-14 \Gamma$) & 64 & 25.4 & Focused to $\unit[12]{mm}$ diameter at the oven exit.\\
\hline
2D MOT & $-25$ ($-0.8 \Gamma$) & 10.5 & 22.8 & 4 beams with centers at the same height as the Zeeman slower axis.\\
\hline
Plug beams & $-13$ ($-0.4 \Gamma$) & 0.017 & 9.6 & 2 symmetric beams inclined at $\unit[8]{^\circ}$ to the vertical ($y$ axis) and aimed at the top half of the blue 2D MOT.\\
		\end{tabular}
		\end{ruledtabular}
\end{table*}

\begin{table*}
\caption{Properties of laser beams used for Red MOT I. Under ``detuning" $\Delta_1 : \delta : \Delta_2$ refers to a modulated comb of lines from $\Delta_1$ to $\Delta_2$ with a spacing of $\delta$.\footnote[1]{Note that due to nonlinearities in the amplifiers and AOMs, it is not possible to separately measure the optical powers in each component of the spectrum.}}
\label{tab:LaserParameters1}
\begin{ruledtabular}
\begin{tabular}{C{0.13\textwidth}C{0.16\textwidth}C{0.1\textwidth}C{0.1\textwidth}C{0.3\textwidth}}
\textbf{Beam name} & \textbf{Detuning [MHz]} & \textbf{Power [mW]} & \textbf{$1/e^2$ diameter [mm]} & \textbf{Comments}\\
\noalign{\smallskip}\hline\hline\noalign{\smallskip}
2D Red molasses & -0.03 : 0.025 : -0.75 & 4.6 & $45.6 \times 18.2$ & 2 sets of 2 counter-propagating beams, centered $\unit[38]{mm}$ below Zeeman slower plane. \\
\hline
Red MOT I Y & -0.95 : 0.017 : -5 & 10.8 & 68 & Waist given $\unit[22]{cm}$ below the quadrupole center. Beam focused on the bottom baffle $\unit[22]{cm}$ above the quadrupole center at $z=\unit[+10]{mm}$ from the central axis of the falling atomic beam.\\
\hline \noalign{\smallskip}
Red MOT I X & -0.7 : 0.015 : -3 & 3.3 & 47 & 2 counter-propagating beams \\
\noalign{\smallskip} \hline
Red MOT I Z Outer & -0.7 : 0.016 : -3 & 1.14 & 48 & 2 counter-propagating beams with an 8mm hole to allow Red MOT I Z Inner beams to pass. \\
\hline \noalign{\smallskip}
\multirow{2}{*}{\parbox[t]{0.13\textwidth}{Red MOT I Z Inner}} & -0.73 : 0.017 : -1.1 & \multirow{2}{*}{0.16\footnotemark[1]}  & \multirow{2}{*}{8} & \multirow{2}{*}{2 counter-propagating beams} \\
			 &   -1.1 : 0.019 : -3.5 &  & & \\
		\end{tabular}
		\end{ruledtabular}
\end{table*}

\begin{table*}
\caption{Properties of laser beams used in the apparatus for Red MOT II(a). Under ``detuning" $\Delta_1 : \delta : \Delta_2$ refers to a modulated comb of lines from $\Delta_1$ to $\Delta_2$ with a spacing of $\delta$.}
\label{tab:LaserParameters2}
\begin{ruledtabular}
\begin{tabular}{C{0.13\textwidth}C{0.16\textwidth}C{0.1\textwidth}C{0.1\textwidth}C{0.3\textwidth}}
\textbf{Beam name} & \textbf{Detuning [MHz]} & \textbf{Power [mW]} & \textbf{$1/e^2$ diameter [mm]} & \textbf{Comments}\\
\noalign{\smallskip}\hline\hline\noalign{\smallskip}
2D Red molasses & -0.03 : 0.025 : -0.75 & 4.6 & $45.6 \times 18.2$ & 2 sets of 2 counter-propagating beams, centered $\unit[38]{mm}$ below Zeeman slower plane. \\
\hline
Red MOT I Y & -0.5 : 0.017 : -4 & 10.8 & 68 & Waist given $\unit[22]{cm}$ below the quadrupole center. Beam focused on the bottom baffle $\unit[22]{cm}$ above the quadrupole center at $z=\unit[+10]{mm}$ from the central axis of the falling atomic beam.\\
\hline \noalign{\smallskip}
Red MOT I X & -0.7 : 0.015 : -3 & 3.3 & 47 & 2 counter-propagating beams \\
\hline \noalign{\smallskip}
Red MOT I Z Outer & -0.7 : 0.016 : -4 & 1.14 & 48 & 2 counter-propagating beams with an 8mm hole to allow Red MOT II Z Inner beams to pass. \\
\hline \noalign{\smallskip}
\multirow{2}{*}{\parbox[t]{0.13\textwidth}{Red MOT II Z Inner}} & -0.1 : 0.017 : -0.25 & \multirow{2}{*}{0.063\footnotemark[1]} & \multirow{2}{*}{8} & \multirow{2}{*}{2 counter-propagating beams} \\
			   & -0.26 : 0.015 : -2.85 & & &  \\
\noalign{\smallskip} \hline \noalign{\smallskip}
\multirow{2}{*}{Red MOT II Y}  & -0.1 : 0.016 : -0.25 & \multirow{2}{*}{0.43\footnotemark[1]} & \multirow{2}{*}{36} &  \multirow{2}{*}{1 upward-propagating beam}\\
			  & -0.26 : 0.014 : -2.85 & & &   \\
\hline \noalign{\smallskip}
\multirow{2}{*}{Red MOT II X} & -0.1 : 0.017 : -0.25 & \multirow{2}{*}{0.099 \footnotemark[1]} & \multirow{2}{*}{28.8} & \multirow{2}{*}{2 counter-propagating beams}\\
			  & -0.26 : 0.013 : -2.85 & & &  \\ 
		\end{tabular}
		\end{ruledtabular}
\end{table*}

\begin{table*}
\caption{Properties of laser beams used in the apparatus for Red MOT II(b). Under ``detuning" $\Delta_1 : \delta : \Delta_2$ refers to a modulated comb of lines from $\Delta_1$ to $\Delta_2$ with a spacing of $\delta$.}
\label{tab:LaserParameters3}
\begin{ruledtabular}
\begin{tabular}{C{0.13\textwidth}C{0.16\textwidth}C{0.1\textwidth}C{0.1\textwidth}C{0.3\textwidth}}
\textbf{Beam name} & \textbf{Detuning [MHz]} & \textbf{Power [mW]} & \textbf{$1/e^2$ diameter [mm]} & \textbf{Comments}\\
\noalign{\smallskip}\hline\hline\noalign{\smallskip}
2D Red molasses & -0.03 : 0.025 : -0.75 & 4.6 & $45.6 \times 18.2$ & 2 sets of 2 counter-propagating beams, centered $\unit[38]{mm}$ below Zeeman slower plane. \\
\hline
Red MOT I Y & -0.5 : 0.017 : -4 & 10.8 & 68 & Waist given $\unit[22]{cm}$ below the quadrupole center. Beam focused on the bottom baffle $\unit[22]{cm}$ above the quadrupole center at $z=\unit[+10]{mm}$ from the central axis of the falling atomic beam.\\
\hline \noalign{\smallskip}
Red MOT I X & -0.7 : 0.015 : -3 & 3.3 & 47 & 2 counter-propagating beams \\
\hline
Red MOT I Z Outer & -0.7 : 0.016 : -4 & 1.14 & 48 & 2 counter-propagating beams with an 8mm hole to allow Red MOT II Z Inner beams to pass. \\
\hline \noalign{\smallskip}
\parbox[t]{0.13\textwidth}{Red MOT II Z Inner} & -0.04 : 0.017 : -0.2 & 0.016 & 8 & 2 counter-propagating beams \\
\noalign{\smallskip} \hline \noalign{\smallskip}
Red MOT II Y  & -0.04 : 0.016 : -0.2 & 0.09 & 36 &  1 upward-propagating beam\\
\hline \noalign{\smallskip}
Red MOT II X & -0.04 : 0.017 : -0.2 & 0.026 & 28.8 & 2 counter-propagating beams\\
		\end{tabular}
		\end{ruledtabular}
\end{table*}

\section{Velocity distribution from the 2D Blue MOT}
\label{sec:Velocity}
	
The atoms exiting the Zeeman slower enter the 2D MOT region. Due to the broad linewidth of the blue transition, they are rapidly cooled to about $\unit[1]{mK}$ in the radial direction, but are presumably unaffected in the 2D MOT, vertical axis. The velocity distribution along this axis depends mostly on the oven output collimation, the transverse cooling stage, and the transverse spread due to scattering from the Zeeman slower beam. This last contribution amounts to about $5 \times 10^4$ spontaneously emitted photons, which gives an energy increase in the slower transverse direction of $\sim \unit[15]{mK}$, part of which will affect the vertical velocity distribution.

\begin{figure}[tb]
\centering
\includegraphics[width=0.98\columnwidth]{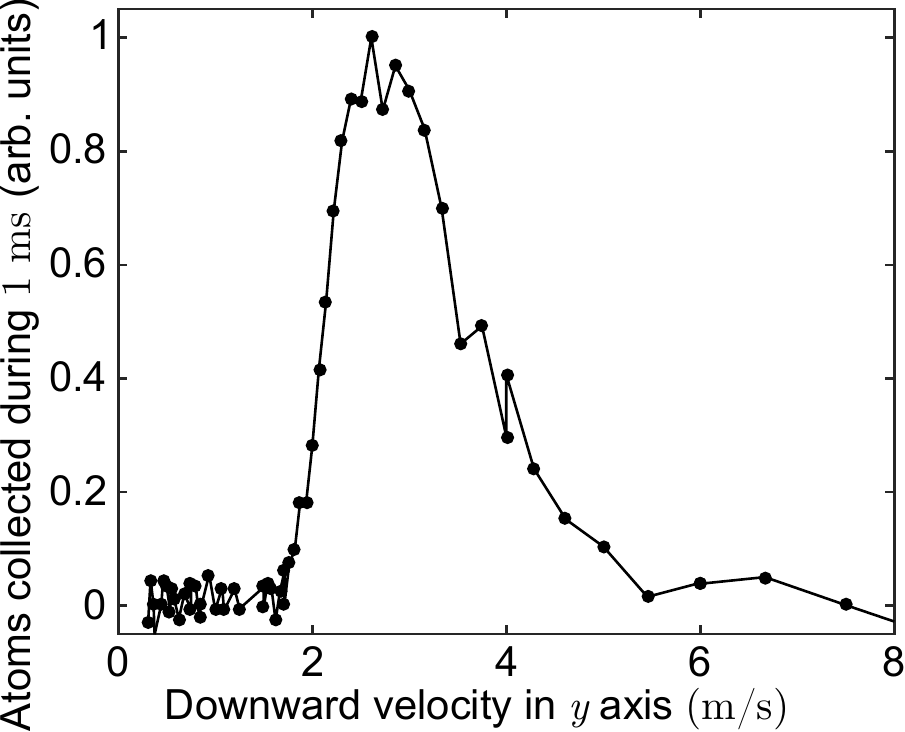}
\caption{\label{fig:SM2_2DMOTVelocityDistribution} Velocity distribution of the atomic beam emitted downward by the blue 2D MOT and captured by Red MOT I, given for atoms located $\unit[55]{mm}$ below the plane of the Zeeman slower.}
\end{figure}

Aiming to measure the vertical velocity distribution, we place two cylindrical $\unit[4]{mm}$ $1/e^2$-diameter high by $\unit[16]{mm}$ $1/e^2$-diameter wide ``gate'' beams in the top chamber, which are actuated by fiber coupled acousto-optic modulator ``switches'' and centered $\unit[40.8]{mm}$ and $\unit[70.8]{mm}$ below the plane of the Zeeman slower. Each beam uses approximately $\unit[2]{mW}$ of light resonant with the blue transition to ``blow away'' any atoms attempting to travel to the bottom chamber. By sequentially opening these blocking beams a known velocity class can be selected. We find that opening any one beam for less than $\unit[5]{ms}$ (with the other beam off) results in no atoms transmitted to the lower chamber, so we use a $\unit[6]{ms}$ gate opening time resulting in an effective opening time of $\unit[1]{ms}$. Atoms are collected in Red MOT I and then compressed by ramping up the gradient and decreasing the detuning to give a cloud dense enough to reliably measure the atom number using absorption imaging. By scanning the time between the opening of the two gate beams, we measure the velocity distribution of the atomic beam emanating from the top chamber that is capturable by Red MOT I, see Figure~\ref{fig:SM2_2DMOTVelocityDistribution}.

\section{Heating from blue MOT photons}
\label{sec:Heating}

We now estimate the heating of Red MOT I by the scattering of fluorescence photons from the 2D blue MOT. The minimum scattering rate from atoms in a red MOT that is required to counter acceleration due to gravity is $\gamma_{\mathrm{min}}=m g/\hbar k$, where $m$ is the atom mass, $\hbar$ is the reduced Planck constant and $k$ is the photon wavenumber. Considering a blue transition photon scattering rate of $\unit[20]{\%}$ of $\gamma_{\mathrm{min}}$ as acceptable, we can estimate the maximum allowable intensity of the blue photon flux at the location of the red MOT, $I_{\mathrm{max}} \approx 0.2 \times 2 \gamma_{\mathrm{min}} I_{\mathrm{sat,blue}}/\Gamma_{\mathrm{blue}} \approx \unit[0.13]{\mu W/cm^2}$, with $I_{\mathrm{sat,blue}}$ and $\Gamma_{\mathrm{blue}}$ being the saturation intensity and the natural linewidth of the blue transition, respectively. Our blue MOT with optical frequency $\nu_{\mathrm{blue}}$ has a flux of $F \approx \unit[2.5 \times 10^9]{atoms / s}$ and atoms spend around $t_{\mathrm{blue}} \approx \unit[20]{ms}$ passing through the blue MOT. The intensity of the blue MOT light is $I_{\mathrm{MOT}} \approx I_{\mathrm{sat,blue}} / 4$. The power scattered by the 2D blue MOT is thus $ P = F t_{\mathrm{blue}} \, h \nu_{\mathrm{blue}} \, \Gamma_{\mathrm{blue}}/2 \, I_{\mathrm{MOT}}/ (I_{\mathrm{sat,blue}} (1+I_{\mathrm{MOT}}/I_{\mathrm{sat,blue}}) )  \approx  \unit[4.1 \times 10^{-4}]{W}$. If we ignore all other sources of blue transition photons such as scattering off the chamber surfaces and windows, and consider only scattered light from the 2D MOT atoms, we need a minimum distance $r$ of $r > \sqrt{P/(4 \pi I_{\mathrm{max}})} \approx \unit[16]{cm}$ to reduce heating in the red MOT to an acceptable level. For the production of a degenerate gas where we require a scattering rate below $\unit[1]{Hz}$ for a reasonable lifetime, the corresponding distance is $ r > \unit[2.8]{m}$. This effectively requires there be no line of sight between a BEC and the 2D blue MOT, or any surface scattering light from the blue MOT, which dictates our apparatus design.

\section{Numerical model results}
\label{sec:Numerical}

\begin{figure*}[tb]
\centering
\subfigure{\includegraphics[width=.60\textwidth]{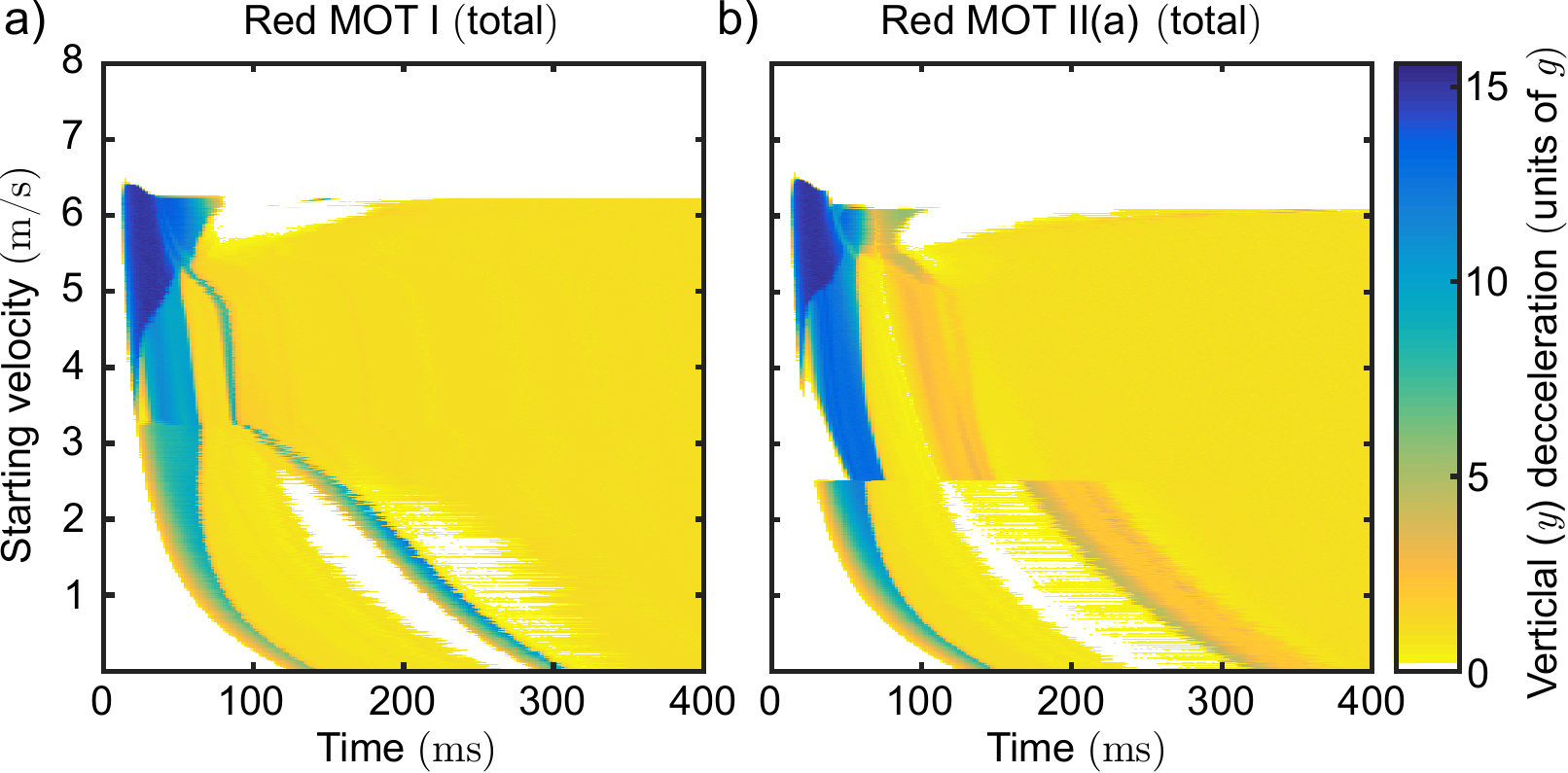}}
\subfigure{\includegraphics[width=.60\textwidth]{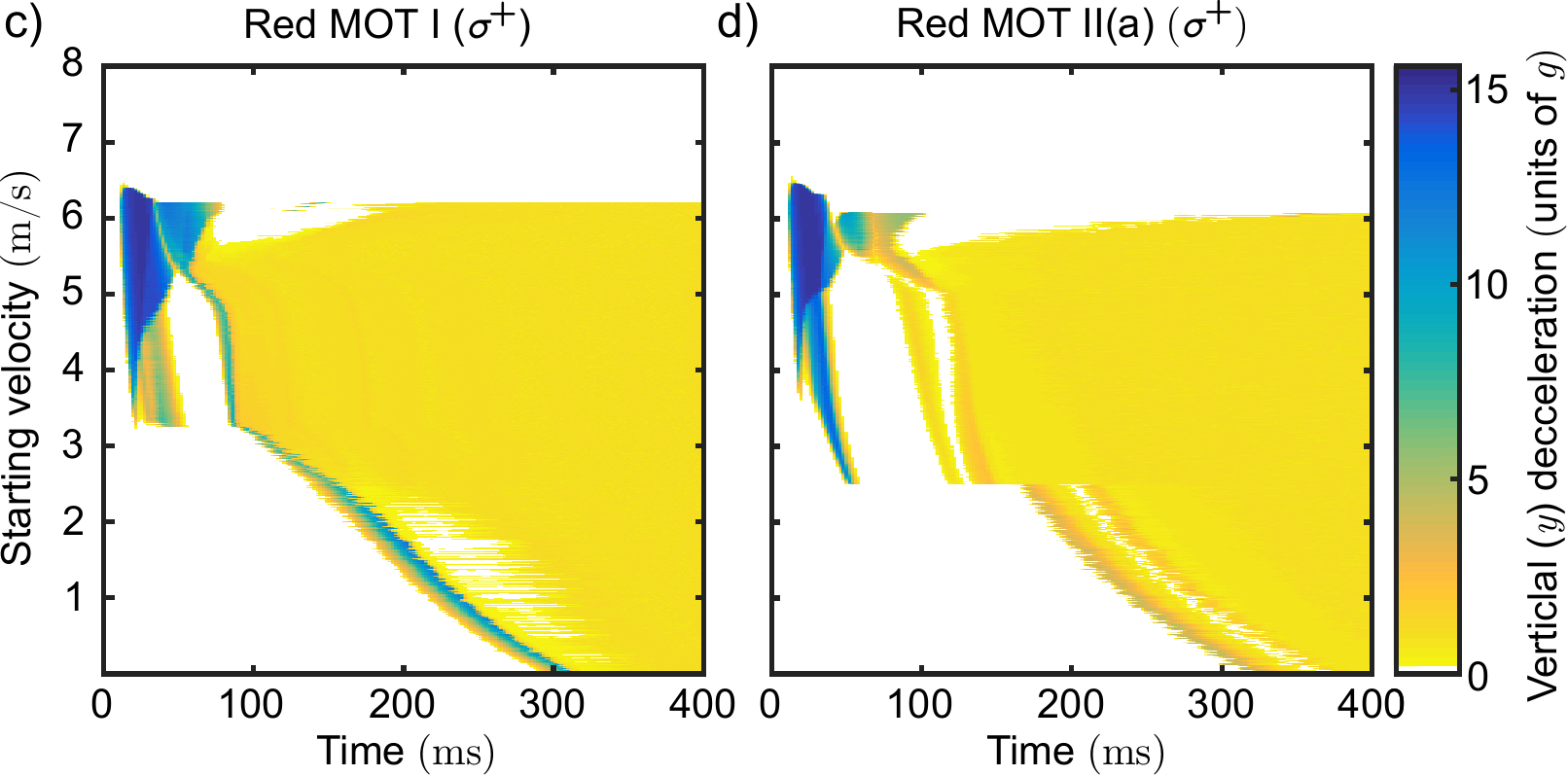}}
\subfigure{\includegraphics[width=.60\textwidth]{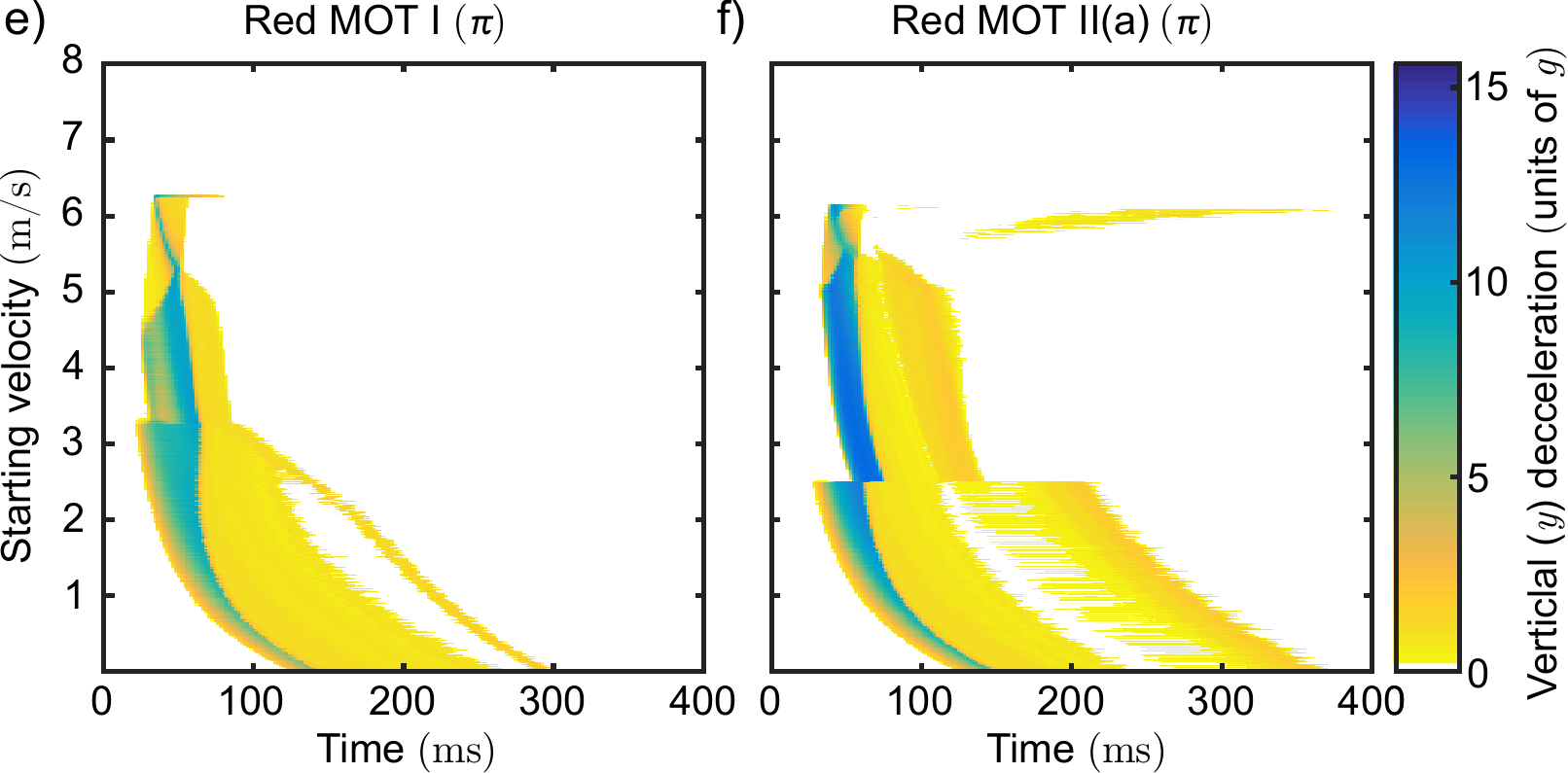}}
\caption{\label{fig:Fig3_HybridSlowerMOTModelling} Monte Carlo numerical simulations of the hybrid slower+MOT. In these simulations, atoms start to fall from the center of the bottom baffle between the chambers, approximately $\unit[20]{cm}$ above the quadrupole center in the case of Red MOT I, with only a downward, $y$ velocity component, which is indicated on the vertical axis. The horizontal axis corresponds to time during 1D trajectories along $y$. (left) Scattering in the Red MOT I configuration and (right) scattering in the Red MOT II(a) configuration. (a-b) Total scattering rate (in units of gravity $g$), (c-d) scattering from the $\sigma^+$ (MOT and Zeeman) and (e-f) scattering from the $\pi$ (non magnetic) components separated. The white regions show decelerations smaller than $0.25 g$.}
\end{figure*}

We now describe the behavior of the hybrid slower+MOT setup, based on our numerical simulations. The results for the 1D idealized situation have already been presented in Figure~2b of the Letter and the 3D Monte Carlo simulations are shown in Figure~2c of the Letter and Figure~\ref{fig:Fig3_HybridSlowerMOTModelling}. The 3D simulations begin with atoms released from the center of the aperture in the bottom baffle approximately $\unit[20]{cm}$ above the quadrupole field center in the case of Red MOT I, and with only a $y$ velocity component. The atoms are then allowed to fall under gravity and experience radiation pressure from the various laser beams and heating from spontaneous emission. Spontaneous emission heating is modeled as a lumped impulse in a random direction proportional to the root of the net radiation impulse absorbed.

In the idealized case, for a low starting velocity $\unit[20]{cm}$ above the quadrupole field center (below $\unit[5]{m/s}$), the falling atoms first experience a Zeeman slower-like deceleration on the $\sigma^+$ transition with a scattering rate close to $\Gamma_{\mathrm{red}} / 2$, where $\Gamma_{\mathrm{red}}$ is the natural linewidth of the red ${^1\mathrm{S}_0} - {^3\mathrm{P}_1}$ transition. This scattering rate corresponds to a deceleration of about $16 \, g$ (blue-colored region in Figure~2b of the letter). As atoms keep decelerating, the Doppler shift $\delta_{\mathrm{Doppler}}$ diminishes and a combination of the Zeeman effect and the broadening $\Delta \nu_{\mathrm{L}}$ of the laser beam frequency dictates the scattering rate at each spatial location. Eventually atoms are almost at rest in the vertical direction and $\delta_{\mathrm{Doppler}}$ becomes negligible. Atoms thus behave like in a standard broadband MOT, and they reach the height where the radiation pressure compensates exactly gravity, typically at the top of the spatial region addressed by the broadband MOT (yellow-colored region labeled ``MOT" in Figure~2b and c of the letter). These atoms are thus captured in the MOT.

For medium starting velocities (between $5$ and $\unit[6.2]{m/s}$ in the idealized case) the atoms' scattering is saturated for most of the time, until atoms are stopped in the vertical direction. This happens at a location deep within the region addressed by the broadband MOT. As a consequence, the scattering rate is way more than what is necessary to compensate gravity, and the hybrid setup acts as a spring and makes the atoms bounce up (see region labeled ``Bounce" in Figure~2b and c of the letter). The relatively weak gravitational force is the only mechanism to push them downward. The simulations indicate that most atoms having bounced will then fall back into the MOT region and get captured there, provided they do not reach surfaces of the vacuum chamber. The horizontal confinement provided by Red MOT I X and Z beams is instrumental to this purpose.

For high starting velocities (above $\unit[6.2]{m/s}$ in the idealized case), the saturated scattering is insufficient to provide enough deceleration for the atoms before they reach the bottom of the spatial region addressed by the broadband MOT, and they are not captured (region labeled ``Fall" in Figure~2b and c of the letter).

The main difference between the idealized case and the more realistic 3D simulation results is the influence of the $\pi$ transition. Due to the tilted magnetic field orientation with respect to the hybrid vertical slower beam, a large part of the interaction occurs on the non-magnetic $\pi$ transition and the $\sigma^+$ Zeeman slower-like transition is less important, as shown in Figure~\ref{fig:Fig3_HybridSlowerMOTModelling}. Moreover, two main extra features appear in the results of the 3D simulations. The first is the disappearance of scattering for starting velocities above $\unit[6.5]{m/s}$, which limits the maximum capture velocity. This originates from the Red MOT I vertical beam frequency boundaries, in combination with its location. Indeed, because of the tilt we apply to the beam in order to hit the side of the bottom baffle, the fastest atoms enter the beam location only after having passed the spatial region where the light is on resonance with the slowing transitions, and thus they are never decelerated. The second feature is the kink in scattering rate appearing for starting velocities of $\unit[3.2]{m/s}$ for Red MOT I and $\unit[2.5]{m/s}$ for Red MOT II(a). This arises when atoms enter the vertical slower beam with a velocity smaller than the minimum velocity addressed by the $\sigma^+$ transition but within the frequency range of the $\pi$ transition. In this case atoms are slowed entirely by the $\pi$ transition before falling at the maximum velocity allowed by the $\pi$ transition into the MOT region.

\end{document}